\documentclass[fleqn,usenatbib]{mnras}

\usepackage{newtxtext,newtxmath}
\usepackage[T1]{fontenc}

\DeclareRobustCommand{\VAN}[3]{#2}
\let\VANthebibliography\thebibliography
\def\thebibliography{\DeclareRobustCommand{\VAN}[3]{##3}\VANthebibliography}

\usepackage{graphicx}	% Including figure files
\usepackage{amsmath}	% Advanced maths commands
\usepackage{subcaption}

\defcitealias{McGill2025}{M25}
\defcitealias{Mackey2019mnras}{M19}
\title[Globular clusters in local group dwarfs]
{Characterising the Globular Cluster Systems of Three Local Group Dwarf Galaxies: NGC~6822, NGC~147 and NGC~185}

\author[G. McGill et al.]{Gracie McGill,$^{1}$\thanks{E-mail: g.r.mcgill@sms.ed.ac.uk (GM)}
Annette M. N. Ferguson,$^{1}$
Dougal Mackey,$^{2}$
Chiara Crociati,$^{1}$ 
Jess M. Howell,$^{1}$
\newauthor
Avon P. Huxor,$^{3}$
Mike J. Irwin,$^{4}$
Alan W. McConnachie,$^{5}$
Alessandro Savino,$^{6}$
Nial R. Tanvir,$^{7}$
\newauthor
Daniel R. Weisz$^{6}$ 
\\
$^{1}$Institute for Astronomy, University of Edinburgh, Royal Observatory, Blackford Hill, Edinburgh, EH9 3HJ, UK\\
$^{2}$Independent Researcher, Charnwood, Canberra, ACT 2615, Australia\\
$^{3}$Department of Computer Science, University of Exeter, North Park Road, Exeter, EX4 4QF, UK\\
$^{4}$Institute of Astronomy, University of Cambridge, Madingley Road, Cambridge CB30HA, UK \\
$^{5}$National Research Council Herzberg Astronomy and Astrophysics, 5071 West Saanich Road, Victoria, BC V9E2E7, Canada \\
$^{6}$ Department of Astronomy, University of California, Berkeley, Berkeley, CA 94720, USA\\
$^{7}$School of Physics and Astronomy, University of Leicester, University Road, Leicester LE1 7RH, UK\\
}

\date{Accepted XXX. Received YYY; in original form ZZZ}

\pubyear{\the\year{}}

\begin{document}
\label{firstpage}
\pagerange{\pageref{firstpage}--\pageref{lastpage}}
\maketitle

% Abstract of the paper
\begin{abstract} % max words = 250 
We present deep HST photometry for 26 globular clusters (GCs) residing in three Local Group dwarf galaxies: the isolated dwarf irregular (dIrr) galaxy, NGC\,6822, and the M31 dwarf elliptical (dE) satellites, NGC\,147 and NGC\,185. From their colour-magnitude diagrams (CMDs), we quantify their red giant branch (RGB) and horizontal branch (HB) morphologies, and employ new empirical relationships to derive measurements of metallicity, reddening and distance.  Additionally, we measure sizes and $V$-band magnitudes from their integrated light profiles.  We find that the clusters span a range of metallicities, from [Fe/H] $\sim-0.7$ to $\lesssim-2$;  however, the three dwarfs have very similar mean GC metallicities ($\sim-1.7$ dex) despite their very different evolutionary histories.  In contrast, we find that almost all of NGC\,6822's GCs exhibit red HB morphologies, whereas those in the two dEs are predominantly blue. We highlight three outlying GCs in NGC\,6822 that have very low metallicities yet very red HBs; they are also the most extended clusters in our sample, with half-light radii of $\sim 13-17$~pc. If these clusters are young, their origin is difficult to explain given their very remote locations, and metallicities which are significantly lower than those of the old and intermediate-age stars in NGC\,6822. These three clusters are strikingly similar to several GCs linked to substructure in the outer halo of M31, suggesting either the recent accretion of an NGC\,6822-like dwarf by M31, or that both M31 and NGC\,6822 have recently accreted a similar low-mass system. Additionally, we find compelling evidence that another NGC\,6822 cluster, SC7, is tidally distorted. 
\end{abstract}

% Select between one and six entries from the list of approved keywords.
% Don't make up new ones.

%%%%%%%%%%%%%%%%%%%%%%%%%%%%%%%%%%%%%%%%%%%%%%%%%%

%%%%%%%%%%%%%%%%% BODY OF PAPER %%%%%%%%%%%%%%%%%%

\section{Introduction}

\begin{figure*}
\hspace*{-2mm}
 \begin{subfigure}{0.335\textwidth}
     \includegraphics[width=\textwidth]{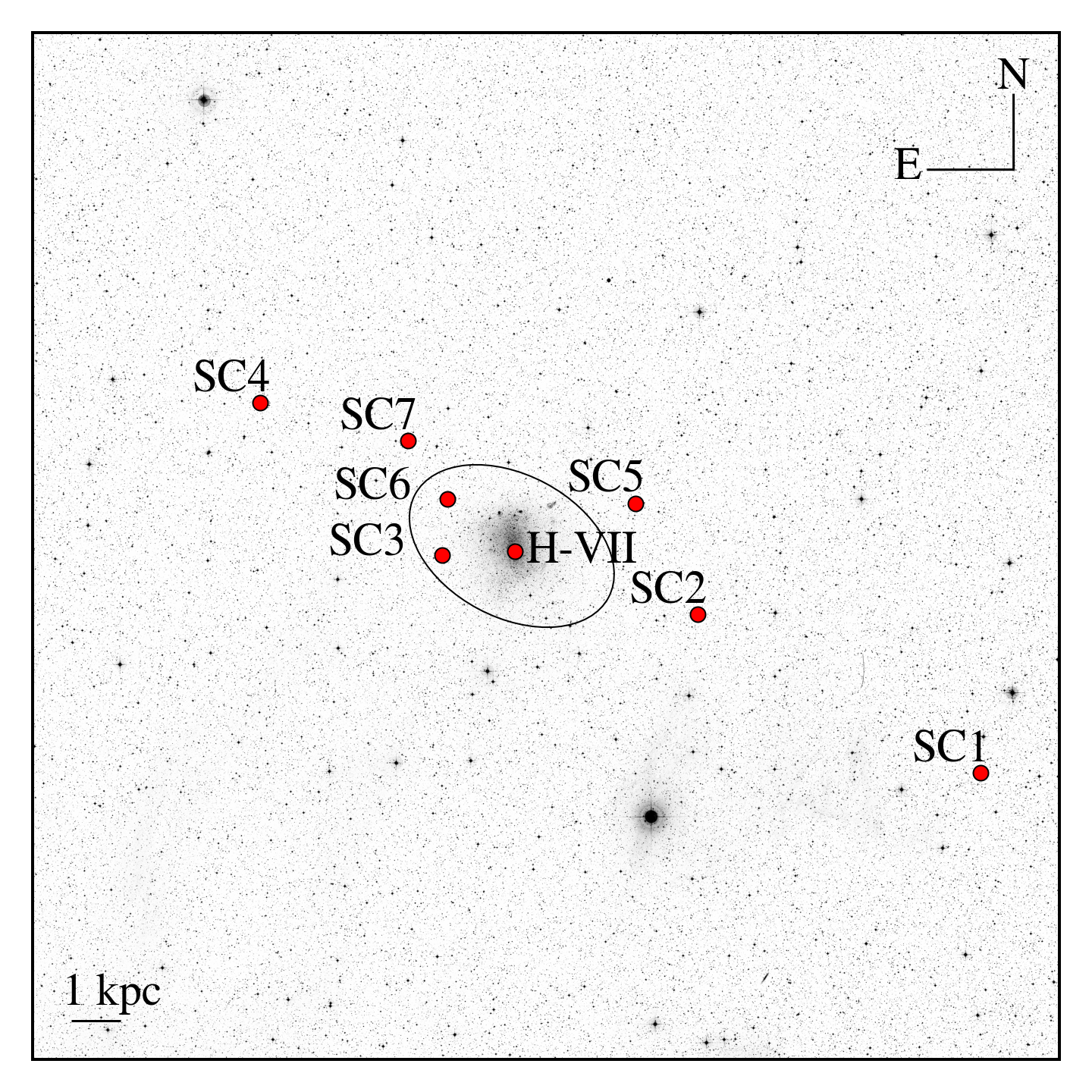}
     \caption{NGC 6822}
     \label{fig:a}
 \end{subfigure}
 \hspace*{-1mm}
 %\hfill
 \begin{subfigure}{0.335\textwidth}
     \includegraphics[width=\textwidth]{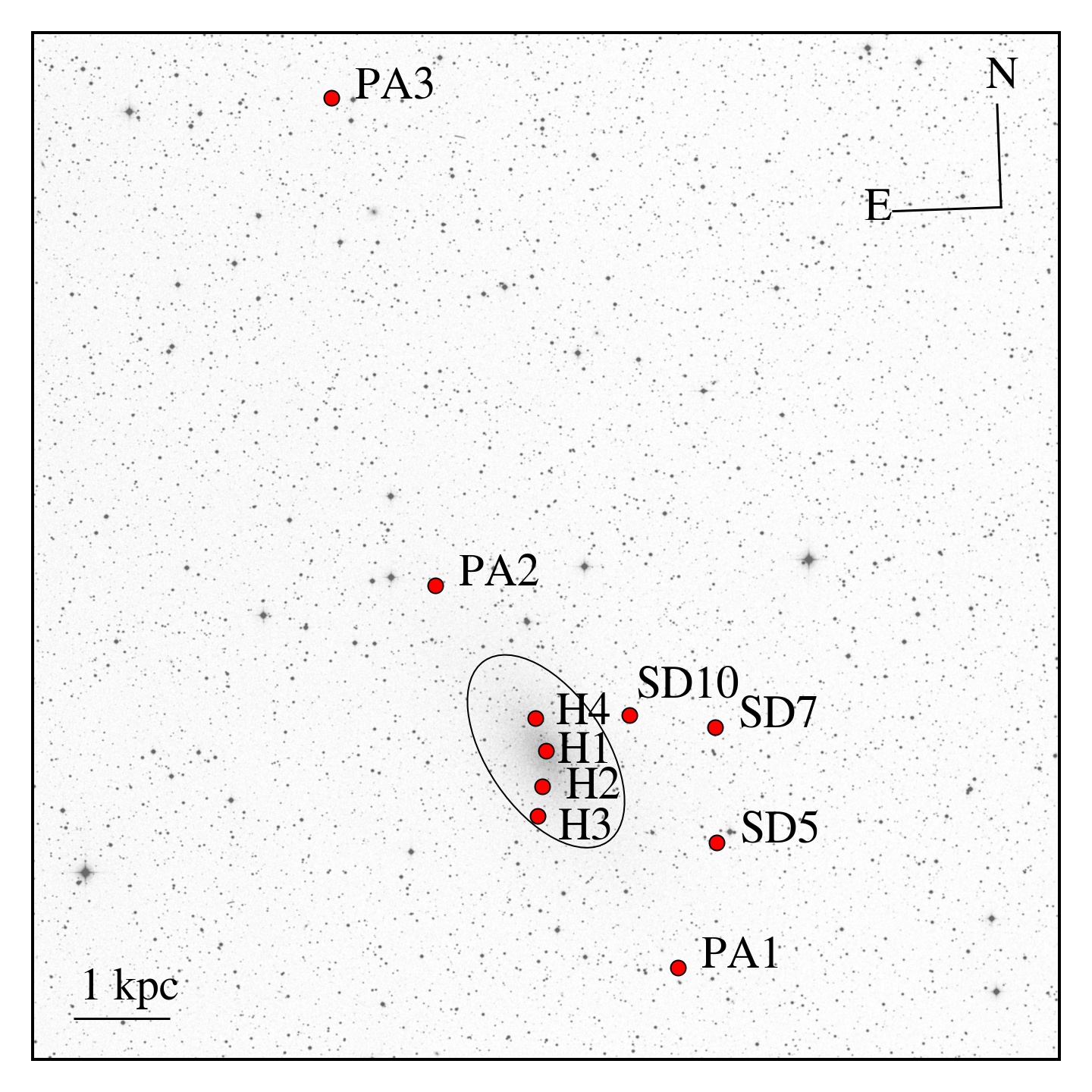}
     \caption{NGC 147}
     \label{fig:b}
 \end{subfigure}
 \hspace*{-1mm}
 \begin{subfigure}{0.335\textwidth}
     \includegraphics[width=\textwidth]{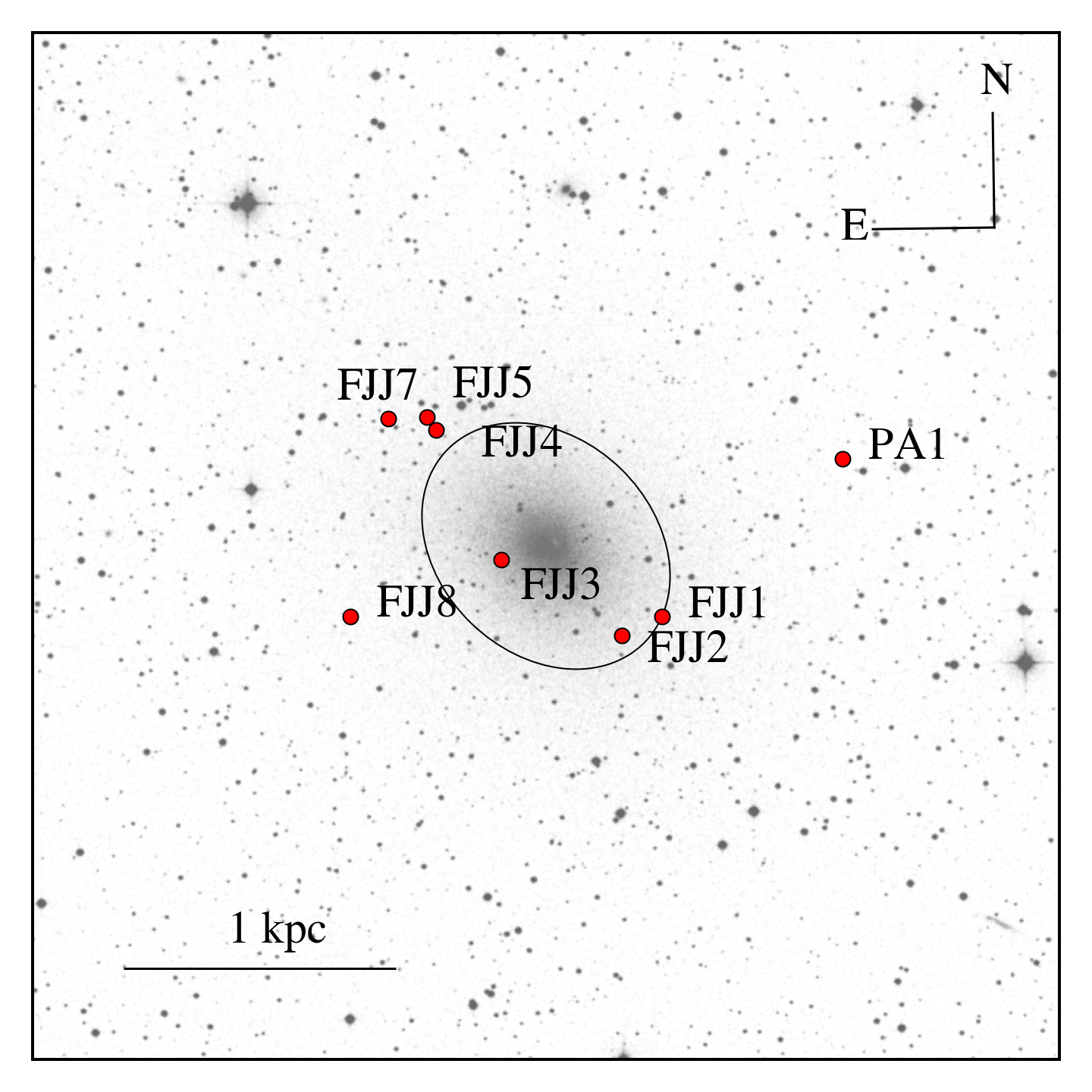}
     \caption{NGC 185}
     \label{fig:c}
 \end{subfigure}
 \caption{
 Positions of GCs in each of the dwarfs, overlaid on Digital Sky Survey images. Half-light ellipses are also indicated \citep{Zhang2021,Crnojevic2014}.
 }
 \label{fig:images}
\end{figure*}

Globular star clusters (GCs) are enigmatic objects that formed in the distant past \citep[e.g.,][]{Adamo2024} and are found in nearly all galaxies with luminosities exceeding a few $10^7$L$_{\sun}$. 
Although the mechanisms and sites of cluster formation remain uncertain \citep[e.g.,][]{Kravtsov2005,Elmegreen2012,Taylor2025,Kruijssen2026}, they are widely recognised as crucial probes of galaxy histories. 

There is growing evidence that not all GCs formed within the potential wells of their current host galaxies, and instead originated in dwarf galaxies that have since been accreted. 
First proposed by \citeauthor{Searle1978} in their seminal 1978 paper, the notion that some fraction of the Milky Way (MW) halo GCs are of external origin has been reinforced via numerous studies of their stellar populations, chemical abundances and kinematics \citep[e.g.,][]{Zinn1993,Mackey2004,Mackey2005,MarinFranch2009,Forbes2010,Myeong2019,Massari2019,Kruijssen2020,Bellazzini2020}. Additionally, many GCs in the outer halo of M31 have also been linked to tidal streams, and hence accretion \citep[e.g.,][]{Mackey2010,Veljanoski2014,Mackey2019mnras,Mackey2019Nat}.
Therefore, understanding the detailed properties of GCs in dwarf galaxies is crucial for gaining insight into the building blocks of the halo GC systems of massive galaxies.

Aside from the Magellanic Clouds, the only two MW satellites to possess sizeable GC systems are the Fornax and Sagittarius dwarf spheroidals. These systems have been studied in considerable detail, and several works have compared their GC populations to the MW halo GC system \citep[e.g.,][]{Mackey2004,   2016AA590A35D,Bellazzini2020,2021ApJ...923...77P}. One commonly used tool in such comparisons is the GC age-metallicity relation (AMR). The AMR has been argued to distinguish between "accreted" and "in-situ" GCs in the MW halo because it shows a clear bifurcation \citep[e.g.,][]{MarinFranch2009, Forbes2020,Horta2021}. It exhibits a steep "in-situ" branch that traces the chemical evolution of the MW itself, and one or more shallower branches populated by "accreted" GCs, which originated in dwarf galaxy environments that enriched more slowly than the MW. Further evidence for this interpretation has recently been presented by \citet{Niederhofer2025}, who used {\it Hubble} Space Telescope (HST) data to study the ages and metallicities of GCs in the Large Magellanic Cloud (LMC). They find that the AMR for LMC GCs is consistent with that of "accreted" GCs in the MW halo, many of which are thought to have originated from the Gaia-Enceladus-Sausage merger \cite[e.g.,][]{Massari2025}.  

In addition to providing insight into the building blocks of more massive galaxies, dwarf GC systems are also of interest because they can reveal the assembly histories of the dwarf galaxies themselves.  \citet{Deason2014} argue that dwarf-dwarf mergers are relatively common and find, in their simulations of the Local Group (LG), that since $z\sim 1$, approximately $10\%$ of satellites more massive than $10^6 M_\odot$ have experienced a major merger. For field dwarfs, this number rises to $15-20\%$. The accreted stellar halos of dwarf galaxies are predicted to be compact and faint \citep{2025A&A...699A..93T, 2025MNRAS.540.2049C, 2025A&A...704A..57C}, making them challenging to detect directly; however GCs may offer a more promising way forward.  Indeed, recent studies have identified compelling evidence of accreted GCs in the LMC \citep{Mucciarelli2021,Niederhofer2025}. On the mass scale of dwarf galaxies, GCs could become invaluable tracers of hierarchical mergers: those deposited by now-defunct satellites may be the only remaining evidence that mergers have occurred. 

In this paper, we present deep HST imaging of the GC populations of three LG dwarfs: NGC~6822, NGC~147, and NGC~185. 
With these data, we present the first sub-horizontal branch (HB) colour-magnitude diagram (CMD) analysis of their GC populations.  We use the CMDs to 
quantify their red giant branch (RGB) and HB morphologies, and employ empirical relationships to derive measurements of cluster metallicity, line-of-sight extinction and distance. 
From their integrated light profiles, we also measure the GC sizes and $V$-band magnitudes.
We discuss how the detailed properties of the dwarf GC systems compare with one another and with the GC populations observed in the MW and M31 halos, finding striking evidence that some of the M31 halo GCs have been recently accreted from an NGC\,6822-like dwarf. 

In Section \ref{background}, we provide a brief overview of the dwarf galaxies whose GC systems we investigate, while Section~\ref{obs} details the observations and data reduction. 
In Section~\ref{mets}, we describe the methodology used to determine metallicity, line-of-sight extinction, and distance from the cluster's CMD and present the resulting measurements.
In Section~\ref{HBmorph}, we quantify the HB morphologies, and in Section~\ref{sizes}, we measure the sizes and luminosities of the clusters via their integrated light profile. 
In Section~\ref{discussions}, we discuss our results, and in Section~\ref{conclusions}, we summarise our key conclusions.

\section{Overview of the target galaxies}\label{background}  
\subsection{NGC 6822}\label{n6822-bkgd}
The dwarf irregular (dIrr), NGC\,6822 \citep[$M_{\star}\sim 3\times10^8 M_\odot$;][]{Pace2025}, is one of the most isolated galaxies within the Local Group, 
and possesses a disc of gas and young stars, which is positioned almost perpendicularly to its main body \citep{deBlok2006}. Although its complex inner structure is suggestive of a recent merger or accretion, no compelling evidence has emerged thus far to support this interpretation. Indeed, the galaxy has a smooth extended stellar spheroid of RGB stars with no signs of tidal streams or other stellar substructure \citep{Battinelli2006, Zhang2021, tantalo2022dwarf-43b}. If NGC\,6822 did experience a past merger, it must have occurred sufficiently long ago for any associated debris to have become fully phase-mixed.  It is suspected that NGC\,6822 may have experienced a weak interaction with the MW, passing within its virial radius $\sim3-4$~Gyrs ago \citep{Zhang2021,Battaglia2022}. However, this is remains uncertain, and \citet{McConnachie2021} argue that the likelihood of a past interaction with the MW is less than 10 per cent.

Until recently, the total number of GCs known in NGC~6822 was eight \citep{Hubble1925,Hwang2011,Huxor2013}, several of which are very remote. Some of the clusters have extended sizes and appear, in projection, to form a linear structure aligned with the semi-major axis of the old spheroid population.  
A recent study based on Early Release Observations (ERO) from {\it Euclid} identified 30 new star clusters in NGC\,6822, including three possible additional faint GCs \citep{Howell2025}. These objects were classified as old ($\geq 5$ Gyrs) and metal-poor ($Z/Z_{\odot} \leq 0.2$) based on fits to their integrated-light spectral energy distributions (SEDs), although CMDs will be required to confirm their nature. The GC population of NGC\,6822 exhibits weak net rotation that follows that of the old stellar population, and its kinematics has been used to infer a dynamical mass of $M_{dyn}\sim 3-4\times10^9 M_\odot$ enclosed within a radius of 11~kpc \citep{Veljanoski2015}. 
High-resolution spectroscopy has been used to measure chemical abundances for three GCs --  SC6, SC7, and Hubble VII -- which have metallicities in the range $-1.1 \lesssim$~[Fe/H]~$\lesssim-1.7$ dex \citep{cohen1998old-83c,Larsen2018,Larsen2022}. Furthermore, whereas SC6 and Hubble VII show enhanced $\alpha$-abundances relative to solar, similar to Galactic GCs, SC7 does not. \citet{Larsen2018} highlight striking chemical similarities between SC7 and Ruprecht 106 (Rup 106), an outer-halo GC in the MW that has previously been suggested to have been accreted \citep{Villanova2013}.  Adding to the unusual nature of SC7 are its high luminosity, large present-day mass \citep[$7.7\times10^5M_{\odot}$,][]{Howell2025} and its elliptical morphology \citep{Huxor2013}.

\subsection{NGC\,147 and NGC\,185}
NGC\,147 \citep[$M_{\star}\sim 1.5\times10^8 M_\odot$;][]{Pace2025} and NGC\,185 \citep[$M_{\star}\sim 1.5\times10^8 M_\odot$;][]{Pace2025} are dwarf elliptical (dE) satellites of M31, currently located at 3D radii of 107 and 154~kpc, respectively.  Due to their proximity to one another, there has been significant debate as to whether they form a bound pair \citep[e.g.][]{vandenbergh1998,Watkins2013}. NGC\,147 exhibits isophotal twisting and tidal tails that are not seen in NGC\,185 \citep{Crnojevic2014}, while NGC\,185 has an older stellar population than NGC~147 by $\sim 4-5$ Gyrs, which might reflect an earlier infall into the M31 halo and, consequently, earlier quenching \citep{Geha2015, Savino2025}. However, NGC\,185 also contains evidence for very recent central star formation, indicating that it has either retained or reacquired cold gas, whereas
NGC\,147 is largely gas-poor. \citet{Arias2016} used a genetic algorithm to explore possible orbital histories for NGC\,147 and NGC\,185, and showed that, despite their disparate present-day properties, their observed features can be reproduced in a scenario in which the galaxies formed a bound pair. In contrast, a more recent analysis of the proper motions of these two galaxies determined that it is very unlikely that they were ever gravitationally bound \citep{Sohn2020}.

NGC\,147 and 185 each host a modest GC system, with most clusters located, in projection, on or near their main stellar bodies \citep{Baade1944,Sharina2009}. A small number of more remote clusters were later discovered in wide-field imaging from the Pan-Andromeda Archaeological Survey \citep{Veljanoski2013}, bringing the total number of GCs in each of the galaxies to 10 and 8, respectively. Detailed chemical abundances from high-resolution spectroscopy have been measured for five GCs in NGC\,147 (PA1, PA2, SD7, Hodge 2 and Hodge 3; $-1.4 \lesssim$ [Fe/H] $\lesssim -2.4$) and two in NGC\,185 (FJJ5, FJJ8; [Fe/H]$\sim -1.75$); all of which were found to be metal-poor and enhanced in $\alpha$-elements relative to solar \citep{Larsen2018,Larsen2022}. 

\section{Observations and Data reduction}\label{obs}
\begin{table}
	\centering
	\caption{Observations.}
    
	\label{tab:obs}
	\begin{tabular}{lccccr} % four columns, alignment for each
		\hline
		  Name & RA & Dec &  Camera & Filter &  Exposure \\
          &&&&& time (s) \\
		\hline
            NGC 6822 \\
            \hline 
            SC1 & 19:40:11.9 & -15:21:46.6 & ACS & F606W & 2337 \\
            &&&& F814W & 2475 \\
            SC2 & 19:43:04.5 & -14:58:21.4 & ACS & F606W & 2331\\
            &&&& F814W & 2469\\
            SC3 & 19:45:40.2 & -14:49:25.8 & ACS & F606W & 2331\\
            &&&& F814W & 2469 \\
            SC4 & 19:47:30.4 & -14:26:49.3 & ACS & F606W & 2308\\
            &&&& F814W & 2469\\
            SC5 & 19:43:42.3 & -14:41:59.7 & ACS & F606W & 2331\\
            &&&& F814W & 2469\\
            SC6 & 19:45:37.0 & -14:41:10.8 & ACS & F606W & 2331\\
            &&&& F814W & 2469\\
            SC7 & 19:46:00.7 & -14:32:35.0 & ACS & F606W & 2331\\
            &&&& F814W & 2469\\
            Hubble-VII & 19:44:55.8 & -14:48:56.2 & ACS & F606W & 2284\\
            &&&& F814W & 2271\\
            \hline
            NGC 147 \\
            \hline
            PA1 & 00:32:35.3 & +48:19:48.0 & ACS & F606W & 2466\\
            &&&& F814W & 2604\\
            PA2 & 00:33:43.3 & +48:38:45.0 & ACS& F606W & 2466\\
            &&&& F814W & 2604\\
            PA3 & 00:34:10.0 & +49:02:39.0 & ACS& F606W & 2466\\
            &&&& F814W & 2604\\
            SD5 & 00:32:22.9 & +48:25:49.0 & WFC3& F606W & 2709\\
            &&&& F814W & 5682\\
            SD7 & 00:32:22.2 & +48:31:27.0 & ACS& F606W & 2460\\
            &&&& F814W & 5148\\
            SD10 & 00:32:47.2 & +48:32:10.7 & ACS & F606W & 4887\\
            &&&& F814W & 10123\\
            Hodge 1 &00:33:12.2 & +48:30:32.3 & WFC3 & F606W & 2709\\
            &&&& F814W & 5682\\
            Hodge 2 & 00:33:13.6 & +48:28:48.7 & WFC3& F606W & 2709\\
            &&&& F814W & 5682\\
            Hodge 3 & 00:33:15.2 & +48:27:23.1 & WFC3& F606W & 2709\\
            &&&& F814W & 5682\\
            Hodge 4 & 00:33:15.0 & +48:32:09.6 & WFC3& F606W & 2709\\
            &&&& F814W & 5682\\
            \hline 
            NGC 185 \\
            \hline 
            PA1 & 00:38:18.8 & +48:22:04.0 & ACS& F606W & 2460\\
            &&&& F814W & 5148 \\
            FJJ1 & 00:38:42.7 & +48:18:40.4 & WFC3& F606W & 2634\\
            &&&& F814W & 5391\\
            FJJ2 & 00:38:48.1 & +48:18:15.9 & WFC3 &F606W & 2634\\
            &&&& F814W & 5391\\
            FJJ3 & 00:39:03.8 & +48:19:57.5 & WFC3 & F606W & 2709 \\
            &&&& F814W & 4116 \\
            FJJ4 & 00:39:12.2 & +48:22:48.2 & ACS& F606W & 2400\\
            &&&& F814W & 4978\\
            FJJ5 & 00:39:13.4 & +48:23:04.9 & ACS& F606W & 2400\\
            &&&& F814W & 4978\\
            FJJ7 & 00:39:18.4 & +48:23:03.6 & ACS& F606W & 2400\\
            &&&& F814W & 4978\\
            FJJ8 & 00:39:23.7 & +48:18:45.1 & ACS& F606W & 2466\\
            &&&& F814W & 2604\\
            \hline
	\end{tabular}
\end{table}
HST observations with the Advanced Camera for Surveys (ACS) and the Wide Field Camera 3 (WFC3) were obtained for a total of 26 GCs (8 in NGC\,6822, 10 in NGC\,147 and 8 in NGC\,185) as part of the Cycle 25 program GO-15336 (PI: Ferguson). These represent the entire populations of confirmed GCs known in these systems\footnote{Our sample excludes the three low luminosity candidates in NGC\,6822 recently discovered by \citet{Howell2025}.}.  Each GC was observed via three dithered exposures in both the F606W and F814W passbands, with typical integration times of 2540s and 3870s, respectively. Details of the observations of the full GC sample are given in Table~\ref{tab:obs}.

The positions of the observed clusters relative to their host galaxies are shown in Figure~\ref{fig:images}. They span a wide range of projected radii from the centres of their respective dwarf hosts, extending out to $\sim 10$~kpc in NGC\,6822 and NGC\,147, whereas all known GCs in NGC\,185 lie within $r_{\rm proj} \lesssim 1.2$kpc. In each dwarf galaxy, at least 50 per cent of the GCs are located beyond the half-light radius.

For each cluster, we performed photometry on the images using the \textsc{Dolphot} 2.0 photometry software \citep{Dolphin2000,Dolphot}. 
\textsc{Dolphot} performs point-spread function (PSF) fitting using model PSFs tailored to the relevant camera. We experimented with different input parameters but ultimately adopted those suggested in the \textsc{Dolphot} user manual, with a few minor tweaks in order to optimise photometry for more crowded fields.
The software provides a variety of photometric quality parameters for each detection; we retained only objects classified as stellar, with valid photometry in all input images, $SNR > 5$ in both filters,  $\text{(sharp}_{F606W} + \text{sharp}_{F814W})^2<0.1$, and $\text{(crowd}_{F606W} + \text{crowd}_{F814W})<1.0$ \citep{Weisz2016}. 

Detection completeness was determined using artificial star tests.
We generated $\sim 10^5$ artificial stars per target, broadly following the distributions of high-quality stellar detections on the image and the CMD. These stars were added to the images and then photometered one-by-one using the built-in \textsc{Dolphot} functionality. Stars were considered recovered if their measured positions lay within one pixel of their input positions and if they passed the same photometric quality cuts applied to the real stars. 
We assigned a completeness value to each high-quality stellar detection by calculating the ratio of recovered to input artificial stars with radial positions and CMD locations comparable to those of the real star under consideration. 
Typical 50 per cent completeness levels were found to be $27.4$ mag in F606W and $26.8$ mag in F814W, corresponding to $\sim2$ magnitudes below the flat part of the HB.

Four clusters in our sample (Hubble-VII, Hodge~1, FJJ2 and FJJ3; flagged in Table~\ref{tab:results}) suffered from extreme crowding and high levels of contamination due to their location in the very dense central regions of their host galaxies. The quality of the resulting photometry was not sufficient for subsequent CMD analysis.  For these GCs,  we therefore present only integrated-light measurements of their sizes and luminosities (see Section~\ref{sizes}). 

Using the photometric catalogues, cluster stars were separated from the surrounding field by imposing a radial cut\footnote{For SC7, an elliptical cut was employed to account for its notably elongated shape (see Section~\ref{SC7})}. The radius of this cut was determined on a case-by-case basis, with tighter cuts used for clusters projecting onto denser background fields (which typically lie at smaller projected radii within their host galaxy). The adopted radii range from $4 - 15$ arcsec and are indicated by red circles in the inset panels of Figures~\ref{fig:CMDs-NGC6822} -- \ref{fig:CMDs-NGC185}. In the case of SC6 and SC7, inner cuts (with radii of $3$ and $4$ arcsecs, respectively) were also applied to mitigate the effects of poor stellar detections in their most crowded central regions.  

\section{Metallicities, Reddenings, and Distances}\label{mets}
\subsection{Method}\label{method-met}
\begin{figure*}
\centering
\includegraphics[width=\textwidth]{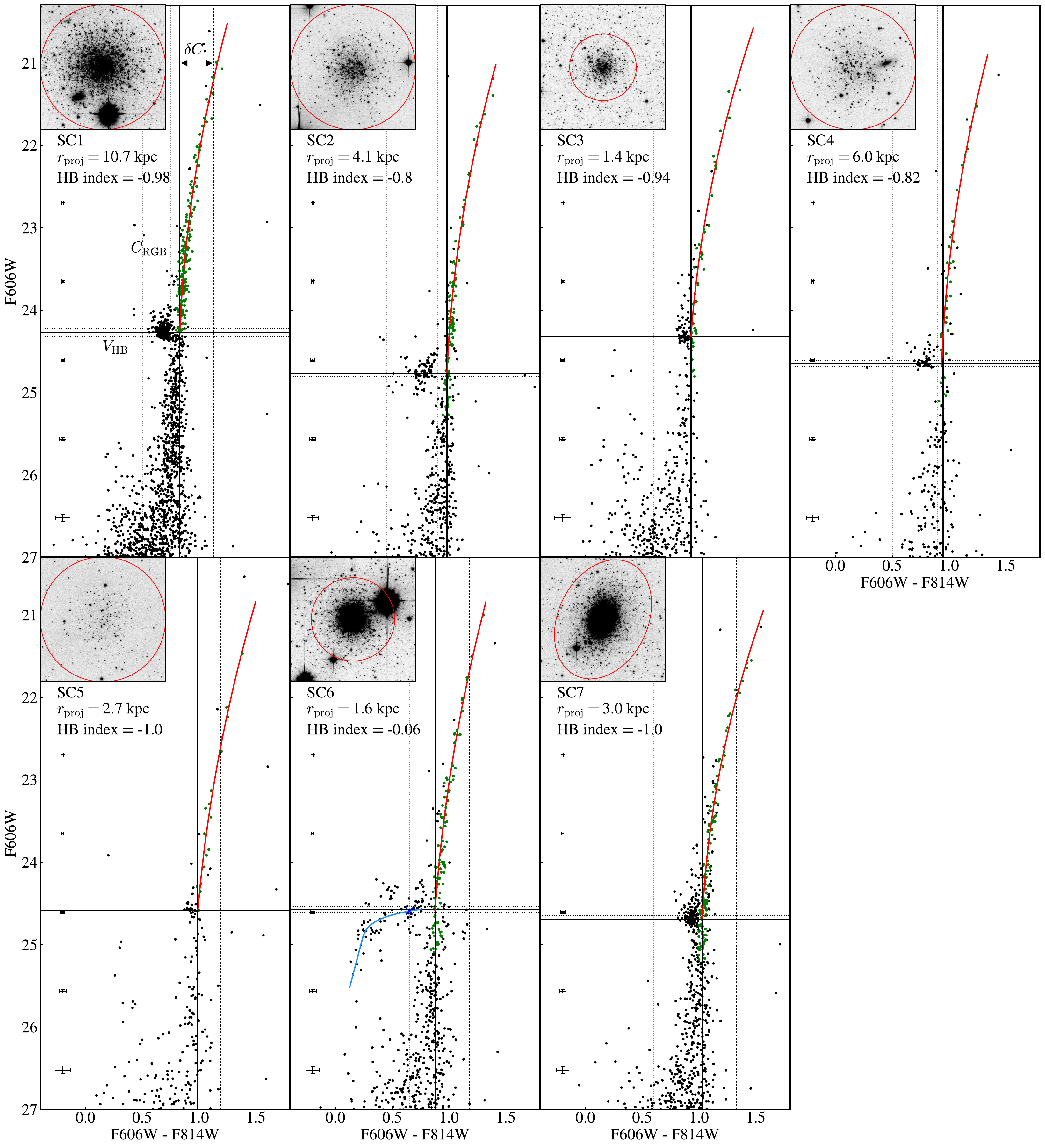}
\caption{CMDs of GCs in NGC\,6822. The inset panels in the top left corner show 30"x30" thumbnails of the GCs in the F606W band, with a red circle indicating the radius within which stars were considered to be cluster members. Below this, the cluster name, projected radius, and HB index are listed. Key measurements are highlighted on the CMDs: the HB magnitude, $V_{HB}$, is indicated by a solid horizontal line; the RGB colour at this magnitude, $C_{RGB}$, is indicated by a solid vertical line; see labels in the top-left panel. For GCs with red HB stars, the range of colour within which $V_{HB}$ was measured is indicated by vertical dotted lines; for SC6, the template HB fit to the blue HB stars is overlaid (blue curve), and the colour at which the HB magnitude is measured is indicated by the vertical dotted line. 
The upper RGB polynomial fit is indicated by the red curve, and the vertical line represents  $\delta C$.}
\label{fig:CMDs-NGC6822}
\end{figure*}

\begin{figure*}
\centering
\includegraphics[width=\textwidth]{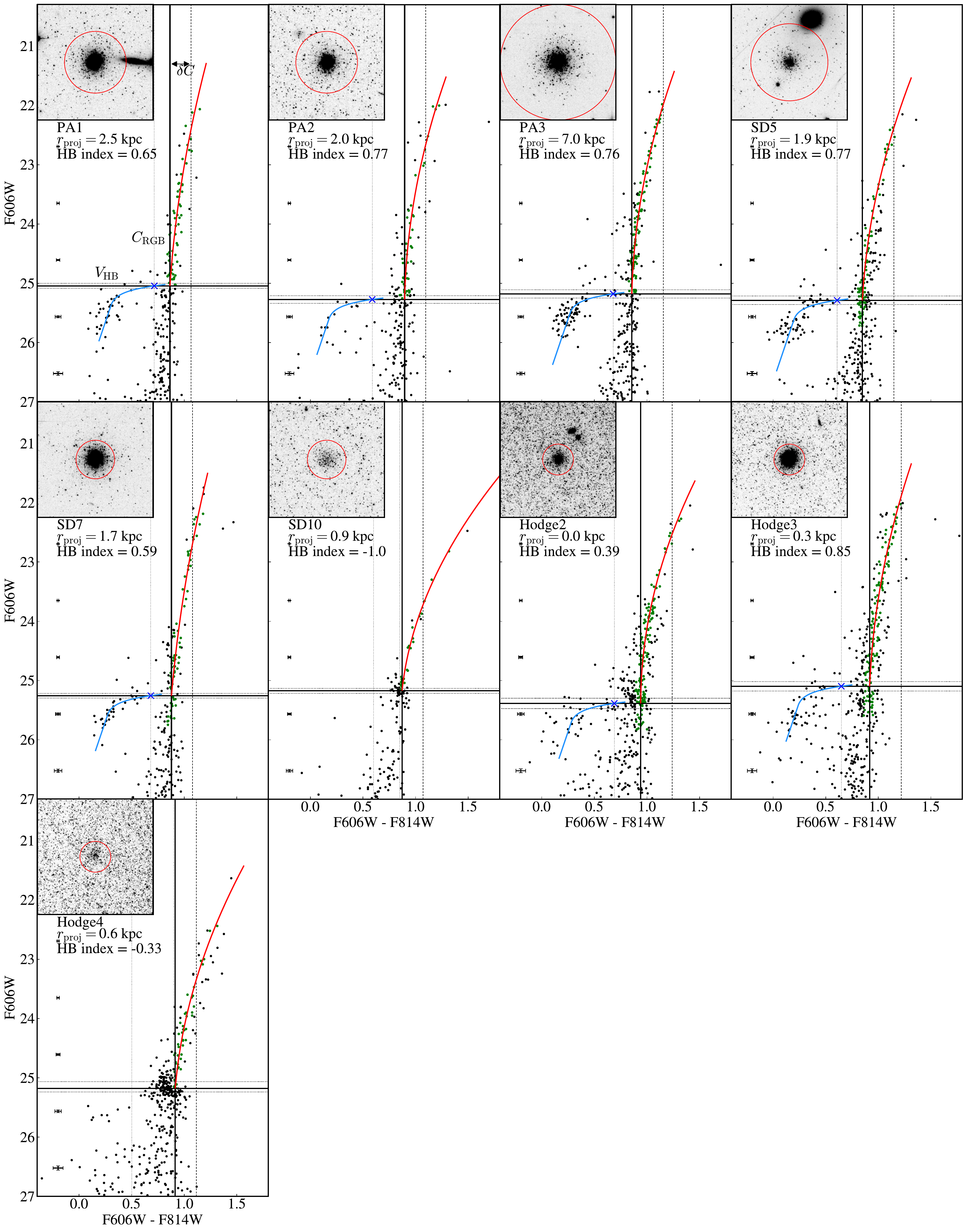}
\caption{CMDs of GCs in NGC\,147, as in Figure~\ref{fig:CMDs-NGC6822}.}
\label{fig:CMDs-NGC147}
\end{figure*}

\begin{figure*}
\centering
\includegraphics[width=\textwidth]{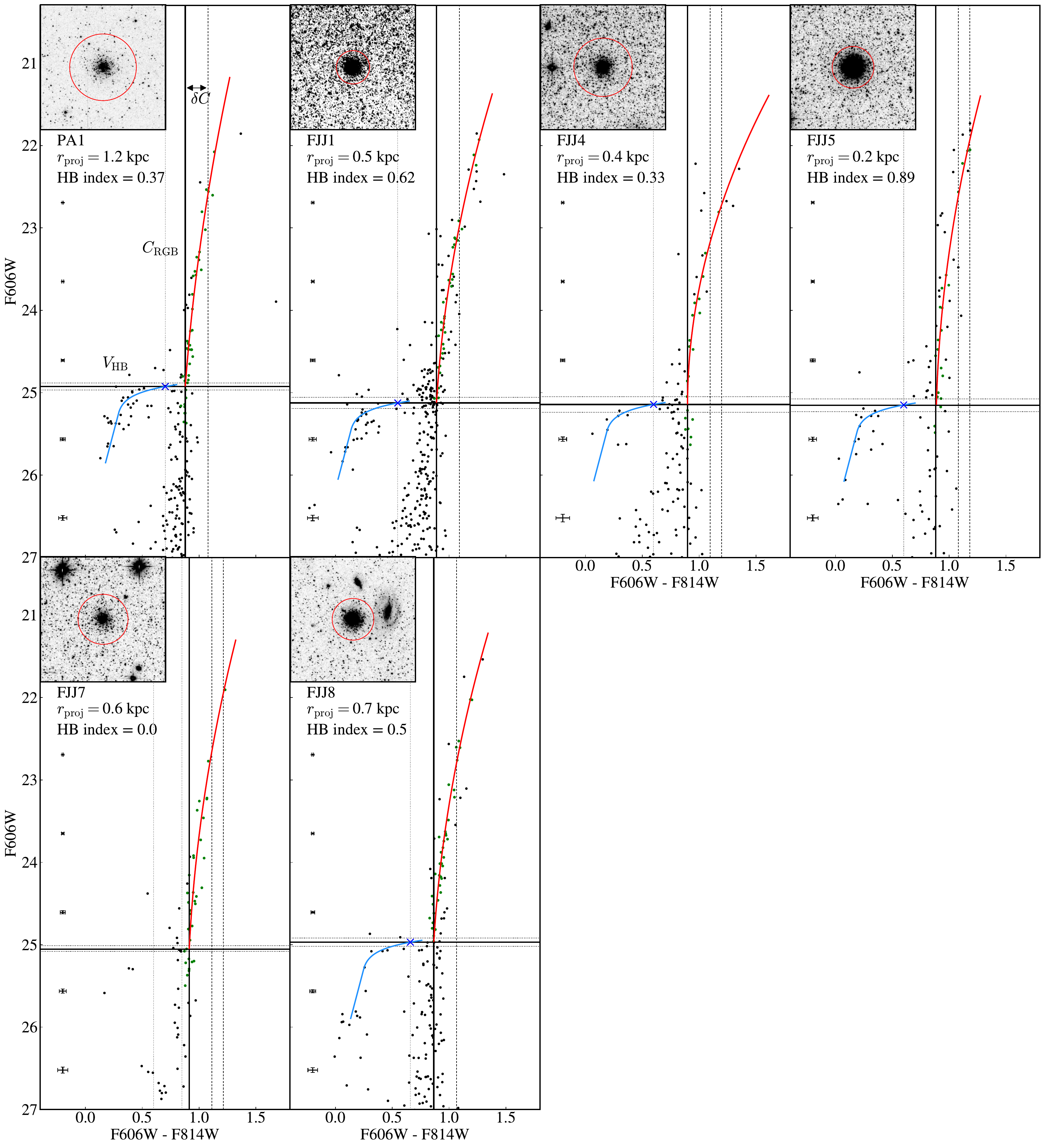}
\caption{CMDs of GCs in NGC\,185, as in Figure~\ref{fig:CMDs-NGC6822}.}
\label{fig:CMDs-NGC185}
\end{figure*}

Figures~\ref{fig:CMDs-NGC6822} -- \ref{fig:CMDs-NGC185} show the CMDs for the sample, as well as 30"x 30" thumbnails of the GCs in the \textit{F606W} band. It can be seen that there is a wide range of CMD morphologies, cluster densities, and concentrations.  These figures also illustrate the methodology used to measure the metallicity, [Fe/H], and the line-of-sight dust extinction, $E(B-V)$. Our approach, which has previously been applied to a sample of 48 outer halo GCs in M31 \citep[hereafter \citetalias{McGill2025}]{McGill2025},
involves the measurement of several indices which are calibrated from empirical relations derived by Mackey et al. (in prep.) using observations of 47 MW GCs from the ACS Globular Cluster Treasury survey \citep[GO-10775,][]{sarajedini2007} and a smaller extension program GO-11586, \citep[GO-11586,][]{Dotter2011}.

The method is inspired by that of \cite{1994AJ....107..618S}, where the key principle is to exploit the dependence of the upper RGB curvature on cluster metallicity. Automated peak-finding algorithms were employed to determine the HB magnitude, $V_{HB}$, and the RGB colour at this magnitude, $C_{RGB}$. A polynomial was then fitted to the upper RGB, anchored at ($C_{RGB}$, $V_{HB}$). The curvature of the upper RGB was quantified as the magnitude difference across a defined colour interval (either $\delta C  = 0.2$ or $\delta C = 0.3$) to the red of this anchor point, which was then used to infer [Fe/H]. 
Given this metallicity, the intrinsic RGB colour could be determined, allowing the line-of-sight reddening, $E(B-V)$, to be measured via comparison to $C_{RGB}$. Finally, once the line-of-sight extinction had been established, we measured the distance to each GC by comparing the observed HB magnitude, $V_{HB}$, with the intrinsic value expected for its metallicity. The resulting measurements are listed in Table~\ref{tab:results}.  Mackey et al. found that, for well-populated RGBs, this method reproduced the spectroscopic metallicities of the MW calibration sample with a precision of $\sim 0.10$ dex when the RGB magnitude difference was measured over a colour interval of $\delta C = 0.3$. Adopting the smaller colour interval of $\delta C = 0.2$ slightly reduces the sensitivity of the RGB curvature to metallicity, yielding a typical precision of $\sim 0.17$. 
They note that the precision of the $E(B-V)$ measurements is more difficult to quantify, because a considerable amount of the scatter arises from the large uncertainties in the MW GC reference values. Nevertheless,  they estimate that the $E(B-V)$ measurements are likely to be approximately $\sim 2-3$ times more precise than those of \citet{Schlegel1998}.

For seven of the GCs in the sample (one in NGC\,6822, four in NGC\,147 and two in NGC\,185), the metallicity determined using the above method was formally below the lower limit of the calibration sample ($-2.5 \la$~[Fe/H]~$\la -0.4$, on the \citet{Carretta2009} scale). There are large uncertainties associated with these metal-poor GCs since the upper RGB slope becomes increasingly insensitive to metallicity in this regime, and since the difference between metal-poor RGB fiducial tracks is comparable to or smaller than the photometric errors. As a result, for these clusters a metallicity floor of $\text{[Fe/H]}= -2.5$ was imposed. The metallicities of these clusters are shown in plots by an upper limit, taken to be the upper 1-$\sigma$ uncertainty associated with the measurement, and are flagged in Table~\ref{tab:results}.

As shown in Figures~\ref{fig:CMDs-NGC6822}--\ref{fig:CMDs-NGC185}, the HB morphology varies significantly across our sample. While some GCs exhibit extended blue HBs typical of old, metal-poor stellar populations (e.g., PA1 in Fig.~\ref{fig:CMDs-NGC147}), others have very red HBs, which may be more consistent with a red clump (e.g., SC7 in Fig.~\ref{fig:CMDs-NGC6822}). This diversity required special care in determining the HB magnitude for a given GC.  To ensure consistency across the full range of HB morphologies, $V_{HB}$ was defined as the magnitude of the HB stars redward of the instability strip. In clusters containing red HB stars, this was determined by a straightforward peak-finding algorithm described above.  However, when no red HB stars are present, this approach is no longer possible. In such cases, a template was fitted to the extended blue portion of the HB, from which the magnitude was measured at the colour where red HB stars would be expected to lie. The template was constructed using six of the sample MW GCs which possess intermediate HB morphologies (Mackey et al., in prep); since the HBs of these clusters were sufficiently populated on both the blue and red sides, both HB-fitting methods could be applied and directly compared to verify consistency. These GCs occupy a relatively narrow metallicity range ($-1.55 \le$~[Fe/H]~$\le -1.45$), since more metal-rich MW GCs within the calibration sample typically have exclusively red HBs, whereas the more metal-poor clusters lack red HB stars. As a result, in cases where it is necessary to apply the `red' HB fitting method to very metal-poor GCs (or the `blue' fitting method to very metal-rich GCs) there may be unknown systematics, since there are no comparably metal-poor red HB GCs are observed in the MW calibration sample (Mackey et al. in prep). Such cases are flagged in Table~\ref{tab:results}.
In addition, for clusters with sparse and/or very broadened blue HBs, the formal uncertainties were increased by a factor of 2 to better capture the uncertainty associated with the blue-HB template fit;  these cases are also flagged in the table. 

For determining the RGB curvature, $\delta C = 0.3$ was the preferred colour interval over which to measure the RGB magnitude, as it provides greater precision in the inferred metallicity. However, if the upper RGB was sparsely populated (e.g., SC5 in Fig.~\ref{fig:CMDs-NGC147}), reliably constraining the fit out to $\delta C = 0.3$ became difficult because of an insufficient number of stars populating the reddest part of the RGB. In such cases,  $\delta C = 0.2$ was instead adopted. 
However, as previously noted, the use of the smaller colour interval results in larger uncertainties, and these cases are also flagged in the table. Specifically, we adopted $\delta C = 0.2$ when $N_{0.3} \leq 5$, where $N_{0.3} $ is defined as the number of stars within 0.5 mag of the RGB magnitude at $\delta C = 0.3$. 

Finally, we also note that the distance for SD10 may be underestimated, as Mackey et al. (in prep.) find that the most metal-rich clusters ([Fe/H]$\gtrsim -1.0$) deviate from the relationship they derive between metallicity and intrinsic HB luminosity. This issue is discussed further in Section~\ref{147-185-dist}.

\begin{table*}
	\centering
	\caption{Results \\
    $^a$ $\delta C = 0.2$ used for estimating metallicity as $N_{0.3} \leq 5$. \\
    $^b$ Derived [Fe/H] $< - 2.5$, so metallicity floor was imposed.  \\
    $^c$ $V_{HB}$ error increased by a factor of 2 to account for uncertainty in blue template fit.\\
    $^d$ Red HB fitting and [Fe/H] $< - 1.6$. \\
    $^e$ [Fe/H] $> - 1.0$, so distance may be underestimated. \\
    $^f$ Severe crowding and field contamination limited the quality of photometry; integrated light measurements only. 
    }
	\label{tab:results}
	\begin{tabular}{lcccccccccr} % four columns, alignment for each
		\hline
		  Name & $R_{proj}$ (kpc) & [Fe/H]  & E(B-V)  & D (kpc) & HB index & $\Delta(V-I)$ & $V$ & $r_h$ (arcsec) &
          $M_V$ & $r_h$ (pc) \\
		\hline
            NGC 6822 \\
            \hline 
            SC1$^d$ & 10.7 & $-2.22\pm ^{0.18}_{0.2}$ & $0.162 \pm^{0.011}_{0.013}$ & $511.1\pm^{16.1}_{16.0}$ & $-0.98 \pm 0.01$ & $0.193 \pm^{0.009}_{0.005}$& 16.23 & 6.4 &$-7.76$ &15.8 \\ 
            SC2$ ^d$ &  4.1 &$-2.1 \pm ^{0.14}_{0.15}$ & $0.321 \pm^{0.008}_{0.009} $ & $531.6 \pm^{12.5}_{11.9}$ & $ -0.8 \pm 0.05$ & $0.269 \pm^{0.012}_{0.01}$ &17.60 & 5.1 & $-6.91$ & 13.3 \\ 
            SC3 & 1.4 &$-1.26\pm^{0.06}_{0.07}$&$ 0.225\pm 0.009$& $447.8\pm^{8.9}_{9.3}$ &$-0.94 \pm 0.06$ & $0.086 \pm ^{0.017}_{0.016}$ &18.40 & 2.9 & $-5.47$ & 6.3 \\
            SC4 $^{a,b,d}$ & 6.0 &$-2.5 \pm^{0.28}_{0.31}$& $0.289\pm^{0.008}_{0.013}$ &$540.1 \pm^{20.6}_{16.0}$ & $-0.82\pm 0.06$ & $0.226 \pm^{0.029} _{0.026}$ &17.93 & 6.3 & $-6.52$ &16.6 \\ 
            SC5 $^a$& 2.7 &$-1.23 \pm^{0.11}_{0.12} $ &$0.287 \pm 0.012$ &$469.6 \pm^{9.8}_{12.5}$ & $-1.0$ & $0.087 \pm^{0.023} _{0.021}$ & 19.59 & 4.3 & $-4.56$ &9.7\\
            SC6 & 1.6 & $-1.74 \pm ^{0.11}_{0.12} $&$0.198 \pm^{0.009}_{0.010}$ & $546.0\pm^{12.6}_{11.7}$ &$ -0.06 \pm 0.08$ & $0.576 \pm^{0.061}_{0.048}$ & 15.94 & 1.3 & $-8.29$ & 3.5 \\ 
            SC7 & 3.0 &$-1.39 \pm ^{0.09}_{0.11}$ & $0.340 \pm 0.010$ &$471.2\pm^{11.6}_{13.7}$&$-1.0$ & $0.121 \pm ^{0.006}_{0.012}$ & 15.39 & 1.3 & $-8.91$ &2.9\\
            Hubble-VII$^f$ & 0.1 &...&...&...&...&...& 16.05 & 0.8 & $-8.11$ & 1.8 \\
            \hline
            NGC 147 \\
            \hline
            PA1$^{a,b}$ & 2.5 & $-2.5\pm^{0.35}_{0.4}$ & $0.2\pm^{0.009}_{0.018}$ &$717.6\pm^{34.6}_{25.1}$ &$0.65 \pm0.1 $ & $0.791 \pm^{0.022} _{0.04}$& 17.12 &  0.9  &$-7.71$ & 3.2 \\
            PA2$^{a,b}$ & 2.0 & $-2.5\pm^{0.74}_{1.0}$ & $0.234\pm^{0.006}_{0.053}$ &$767.5\pm^{100.6}_{44.1}$ &  $0.77 \pm0.13$ & $0.872 \pm^{0.068} _{0.072}$ & 17.46 & 0.9  &$-7.61$ & 3.3 \\
            PA3$^{b,c}$ & 7.0 & $-2.5\pm^{0.34}_{0.4} $& $0.192\pm^{0.011}_{0.019}$ &$770.7\pm^{43.8}_{32.5}$ &  $0.76 \pm 0.06$ & $0.804 \pm^{0.027} _{0.023}$ & 17.57 & 2.5 &$-7.39$& 9.4 \\
            SD5 & 1.9 &$-1.77\pm^{0.22}_{0.28}$ &$0.171\pm^{0.011}_{0.014} $& $777.8\pm^{36.0}_{30.0}$ & $0.77 \pm 0.06$ & $0.718 \pm^{0.021} _{0.041}$& 18.17 & 1.8 &$-6.75$ &6.9 \\
            SD7$^{a,b}$ & 1.7 & $ -2.5\pm ^{0.26}_{0.29} $&$0.215\pm^{0.008}_{0.013}$ &$777.8\pm^{28.7}_{22.4}$ & $ 0.59 \pm 0.11$ & $0.775 \pm^{0.039} _{0.017}$ &17.16 &  1.0  &$-7.88$ &3.8 \\
            SD10$^{a,e}$ & 0.9 & $-0.66\pm^{0.19}_{0.16}$ &$0.107\pm^{0.019}_{0.024}$ &$716.5\pm^{25.8}_{24.8}$ &  $-1.0$ & $0.058 \pm^{0.019} _{0.009}$ & 20.14 & 1.9 & $-4.43$ & 6.5\\
            Hodge 1$^f$& 0.03 &...&...&...&...&...& 17.48 & 1.85 & $-7.32$ & 6.6\\
            Hodge 2$^c$ & 0.3 & $-1.66\pm^{0.37}_{0.6}$ & $0.263\pm^{0.038}_{0.045}$ & $725.4\pm^{65.9}_{44.7}$ &  $0.39\pm 0.11$ & $0.672 \pm^{0.062} _{0.039}$ & 18.31 & 1.2 &$-6.71$ & 4.3 \\ 
            Hodge 3$^{c}$ & 0.6 & $-2.37\pm^{0.36}_{0.45}$ & $0.258\pm^{0.024}_{0.034}$ & $680.8\pm^{48.7}_{35.2}$ &$0.85 \pm 0.05$ & $0.708 \pm^{0.051} _{0.05}$ & 16.72 & 1.4 &$-8.15$ &4.5\\
            Hodge 4$^{a}$ & 0.4 &$-1.07\pm^{0.19}_{0.21}$ & $0.195\pm^{0.019}_{0.020}$ & $676.0\pm^{42.1}_{24.8}$ &  $-0.33 \pm 0.21$ & $0.23 \pm^{0.059} _{0.04}$ & 19.97 & 1.2  & $-4.71$& 3.9 \\
            \hline 
            NGC 185 \\
            \hline 
            PA1$^a$ & 1.2 & $-2.08\pm^{0.31}_{0.36}$ &$0.212\pm^{0.008}_{0.014}$ &$646.3\pm^{28.6}_{21.3}$&  $0.37 \pm 0.12$ & $0.676 \pm^{0.026} _{0.039}$ &19.15 &1.6 &$-5.48$& 5.0 \\
            FJJ1$^{c}$ & 0.5 & $-1.53\pm^{0.17}_{0.22}$ & $0.202\pm 0.013$ & $681.4\pm^{28.3}_{24.8}$ & $0.62 \pm 0.12$ & $0.686 \pm^{0.032} _{0.027}$ &18.39 &1.0& $-6.33$ &3.4 \\
            FJJ2$^f$ &0.2 & ... &...&...&...&...& 18.74 &1.2 & $-5.80$ & 3.7 \\
            FJJ3$^f$ & 0.6 &...&...&...&...&...& 16.65 & 1.6 &$-7.91$ & 4.9  \\
            FJJ4 & 0.6 & $-1.22\pm^{0.21}_{0.31}$ & $0.190\pm^{0.028}_{0.027}$ & $677.7\pm^{42.9}_{37.0}$ & $0.33 \pm 0.31$ & $0.695 \pm^{0.024} _{0.085}$& 18.14 &  0.9 & $-6.54$ & 2.9  \\
            FJJ5$^e$ & 0.7 &$-2.5\pm ^{0.73}_{1.0}$ & $0.22\pm^{0.014}_{0.059}$ & $738.4\pm^{99.6}_{46.3}$ &  $0.89\pm 0.11$ & $0.907 \pm^{0.034} _{0.108}$ &16.82 & 1.2 & $-8.12$ & 4.2 \\
            FJJ7$^d$ & 0.8 &$-2.38\pm^{0.43}_{0.5}$ &$0.252\pm^{0.006}_{0.018}$ &$672.8\pm^{40.6}_{23.2} $ & $0.04\pm 0.41$ & $0.628 \pm^{0.135} _{0.248}$ & 18.88 & 1.4 & $-5.95$ & 4.4 \\ 
            FJJ8$^a$ & 0.8 & $-1.57\pm^{0.19}_{0.22}$ & $0.178\pm^ {0.013}_{0.014}$ & $654.2\pm^{22.9}_{19.9}$ & $0.5 \pm 0.16$ &$0.779 \pm^{0.075} _{0.151}$& 17.72 & 0.9 &$-6.85$ &2.8 \\
            \hline
	\end{tabular}
\end{table*}

\subsection{Comparison to the literature}\label{results}

\begin{figure*}
\centering
\includegraphics[width=\textwidth]{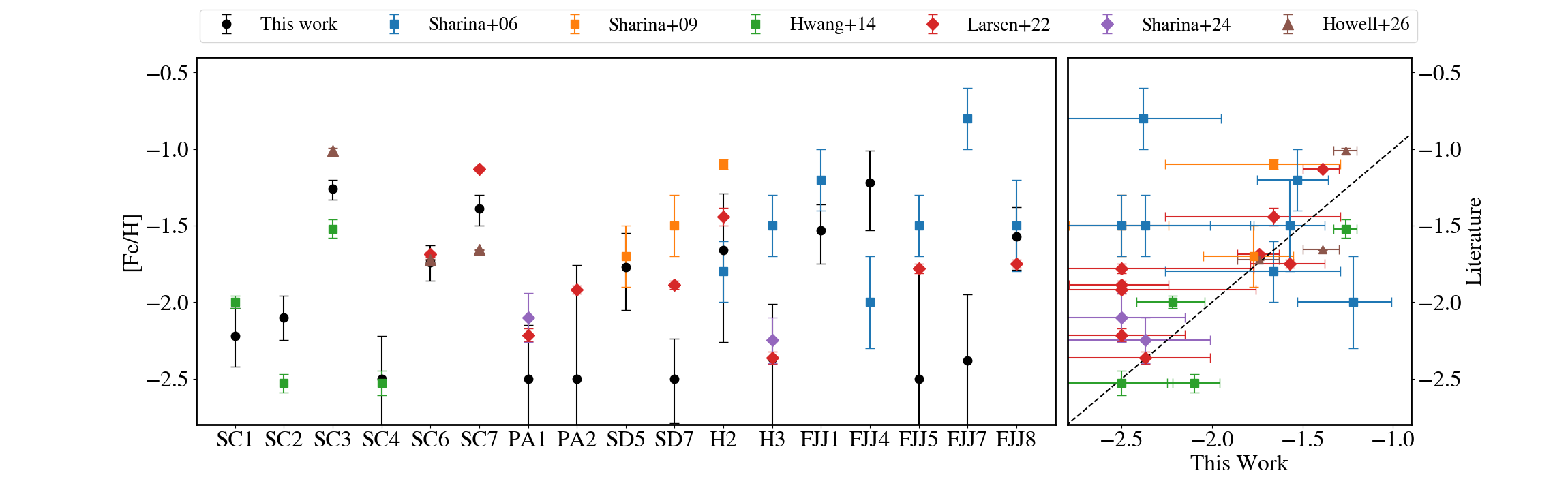}
\caption{Comparison of [Fe/H] with literature studies. The symbol shapes highlight the methodologies employed by these works: squares for Lick Index measurements \citep{Sharina2006,Sharina2009,Hwang2014}, diamonds for spectroscopy \citep{Larsen2022,Sharina2024}, and triangles for SED fitting \citep{Howell2025}. \textit{Left:} Detailed comparison of literature metallicities for each cluster. \textit{Right:} measured metallicity vs literature metallicity.}
\label{fig:met-comp}
\end{figure*}

Table~\ref{tab:results} presents our measurements of [Fe/H], E(B-V) and distances for the GC sample.  Our study presents the largest sample of dwarf galaxy GCs with CMD-based metallicity measurements that have been derived robustly and consistently. Nevertheless, as with all methods of GC metallicity determination, small systematic effects may be present. To investigate this, Figure~\ref{fig:met-comp} compares our CMD-based measurements with values from the literature. The left panel shows the literature values for each cluster, with our measurements overplotted in black, while the right panel compares the literature metallicities directly with our own. Although the figure exhibits considerable scatter, it is important to note that the literature itself already shows significant disagreement, reflecting the wide range of methodologies that have been employed.

\citet{Larsen2022} measured detailed chemical abundances for eight GCs in our sample (two in NGC\,6822, five in NGC\,147, and one in NGC\,185) via high-resolution integrated-light spectroscopy. Compared with their measurements, our CMD-based results are, on average, $\sim 0.2$ dex more metal-poor. However, when the four clusters that reached the imposed metallicity floor are excluded, the mean offset decreases to $\sim 0.1$ dex.  Of these excluded clusters, \citet{Larsen2022} found three to have metallicities of [Fe/H]~$\lesssim -1.9$, thus supporting the conclusion that they are indeed very metal-poor. The exception is FJJ5, for which they measure a metallicity of [Fe/H]$= -1.780 \pm 0.031$. Although this value is substantially more metal-rich than the $-2.5$ floor, it remains consistent with our CMD-based measurement within the uncertainties, which in this case are large due to the sparsity and scatter of the RGB and HB. 

The NGC\,6822 cluster SC7 presents an interesting case for comparison, as it is known to have a solar-scaled alpha abundance \citep{Larsen2018, Larsen2022}. For this cluster, we find that the spectroscopic metallicity is higher \citep[$\lbrack \text{Fe/H}\rbrack=-1.130\pm 0.011$,][]{Larsen2022} than our CMD-based value by $\sim 0.3 $~dex. However, our metallicity determination is calibrated on a sample of MW GCs that are typically $\alpha$-enhanced relative to solar ([$\alpha$/Fe]$\approx +0.4$).  Using BaSTI isochrones \citep{Pietrinferni2021}, Mackey et al. (in prep.) showed that, for a cluster with solar $\alpha$-abundances ([$\alpha$/Fe]~$\approx 0.0$), our method may underestimate the metallicity by as much as $\sim0.3$ dex, entirely consistent with the offset measured for this cluster.

Low-resolution spectroscopic metallicities, as well as metallicities derived from Lick indices, have been reported for several clusters by \citet{Sharina2024} and \citet {Sharina2006,Sharina2009,Hwang2014}, respectively. Although these measurements are known to have considerable uncertainties, they still provide a useful point of comparison for our CMD-based measurements.  \citet{Sharina2024} measured metallicities for two GCs associated with NGC\,147 (PA1 and Hodge 3), and, for both of these clusters, their measurements agree with those presented here within the uncertainties. 
For the seven GCs in NGC\,147 and NGC\,185 with Lick index measurements from \citet{Sharina2006}, there is generally reasonable agreement with the CMD-based values presented here. For the three clusters in their sample that reached the imposed metallicity floor of $-2.5$ in our analysis (Hodge 3, FJJ5 and FJJ7), their values are typically $\gtrsim 1$ dex higher. However, recent high-precision spectroscopic measurements of Hodge 3 \citep{Larsen2022,Sharina2024} confirm that this GC is metal-poor, bolstering confidence in our measurement.  \citet{Sharina2009} estimated ages and metallicities of a further three GCs in NGC\,147 and found all of them to be old and metal-poor. Our result for SD5 is in good agreement with theirs, whereas for Hodge\,2 and SD7, their Lick-index metallicities are higher. This discrepancy is especially notable for SD7, which hits the metallicity floor in our work but for which their estimate is $\sim1$ dex higher. However, for both clusters, spectroscopic metallicities from \citet{Larsen2022} favour lower metallicities than those reported by \citet{Sharina2009}, suggesting that the latter may have overestimated the metallicities. 
\citet{Hwang2014} determine Lick-index metallicities for four GCs in NGC\,6822 GCs, and their values are in good agreement with ours, with a mean difference of only $-0.13$ dex.

More recently, \cite{Howell2025} measured metallicities for GCs in NGC\,6822 using {\it Euclid} ERO and archival ground-based data. These values were derived from SED fits to photometry in the $UBVRIYJH$ bands. Three of the GCs in the present sample were included in their study, and their metallicities (assuming $Z_\odot$ = 0.02) are in good agreement with ours, with a mean offset of only $\sim 0.04$ dex. We note, however, that \cite{Howell2025} obtain a rather young age for SC3 ($\sim2.5$ Gyrs), inconsistent with its previous classification as a GC; they suggest that the SED fitting for this faint cluster is uncertain and its metallicity may be overestimated. Indeed, we find that SC3 is likely older than their estimate, given the absence of a bright main sequence in its CMD. 

\begin{figure}
\centering
\includegraphics[width=\columnwidth]{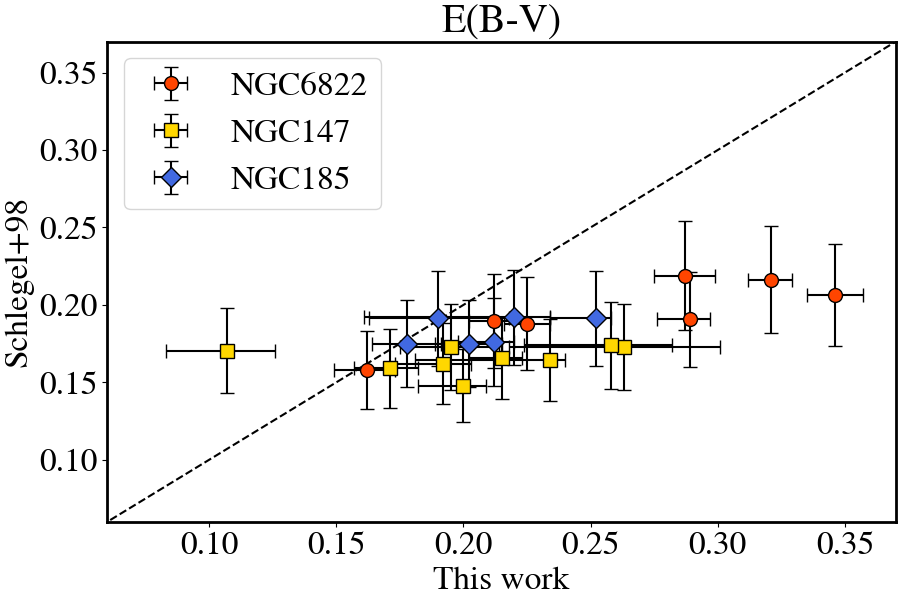}
\caption{Comparison of the $E(B-V)$ values measured in this study with those of \citet{Schlegel1998}, as recalibrated by \citet{Schlafly2011}.}
\label{fig:ebv-res}
\end{figure}

In Figure~\ref{fig:ebv-res}, we compare the measured line-of-sight reddening, $E(B-V)$, with the values from \citet{Schlegel1998}, as recalibrated by \citet{Schlafly2011}. While the mean difference between the measurements is only $\sim 0.04$ mag, our CMD-based measurements are almost always higher. 
For several of the GCs in NGC\,6822 (SC2, SC4, SC5, SC7), the discrepancy is more pronounced, with our measurements in the range $E(B-V)\sim0.3-0.35$ compared to the dust-map values of $E(B-V)\sim0.2$. However, our measurements are consistent with more recent literature estimates for the extinction towards the centre of NGC\,6822 \cite[$E(B-V)\sim 0.35$, e.g.][]{Fusco2012,Rich2014}.  
NGC\,6822 lies at low Galactic latitude ($b\sim-18.4$ degrees) and so it is likely to be affected by spatially variable foreground reddening. Small offsets from the values predicted by the \citet{Schlegel1998} dust maps are thus not unexpected, given their relatively coarse spatial resolution ($\sim 6$ arcmin) compared to the pencil-beam measurements made here, which are more sensitive to small-scale dust structure along the sight-line. Indeed, a recent study by \citet{2026arXiv260111997A} demonstrated how richly structured the foreground dust toward another low Galactic-latitude dwarf galaxy, IC\,10, is relative to the \citet{Schlegel1998} maps (compare their Figs.~3 and~4).

\subsection{Metallicity and the Red Giant Branch Bump}\label{RGBbump}
\begin{figure}
\centering
\includegraphics[width=\columnwidth]{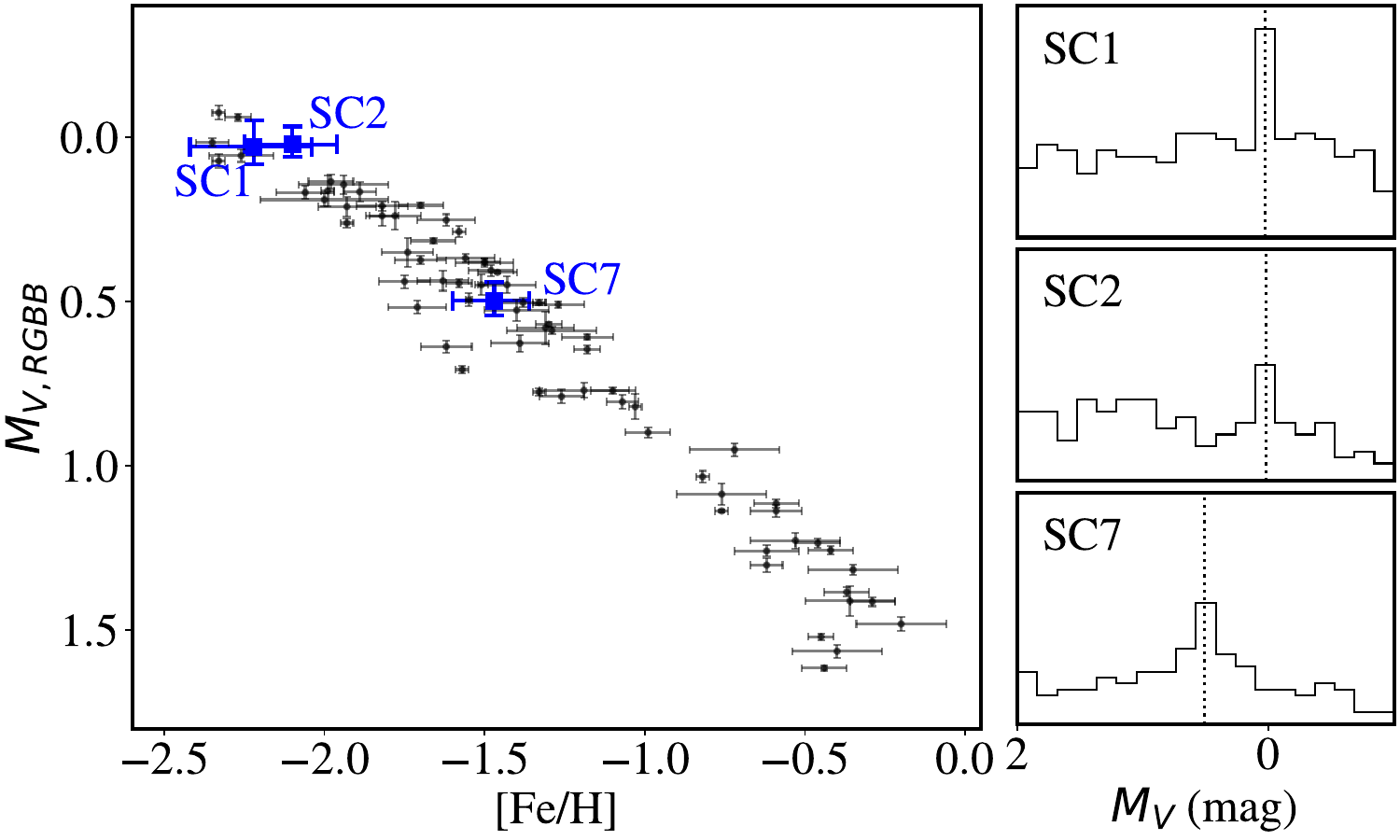}
\caption{\textit{Left:} Absolute $V$-band magnitude of the RGBb against metallicity. Grey points show Galactic GC data from \citet{Nataf2013}; GCs from this work are shown and labelled in blue. \textit{Right:} $V$-band RGB luminosity functions for SC1, SC2 and SC7. The RGBb luminosity is indicated by the dotted line.
}
\label{fig:SC1-rgbb}
\end{figure}

From the CMDs, we note that three of the GCs in NGC\,6822 (SC1, SC2 and SC7) exhibit a weak red giant branch bump (RGBb). The RGBb occurs as a result of the outward-moving H-burning shell crossing the chemical discontinuity at the innermost point reached by the convective layer. The decreased H-abundance slows the rate of fusion, and so the star briefly dims before proceeding up the RGB as before \citep{Sweigart1990,Salaris2002}. As a result, in a sufficiently well-populated RGB, this feature appears as a visible clump of stars on the CMD. 
The brightness of this feature is sensitive to metallicity \citep[e.g.,][]{Salaris2002} and so can be exploited as an independent check on the metallicity of these clusters. 

Comparisons with stellar models are complicated by significant uncertainties. For example, it is well known that theoretical predictions typically underestimate the RGBb brightness by as much as $\sim 0.4$ mag at low-metallicities \citep[e.g.][]{Cassisi2011}. An empirical approach is therefore preferable. \citet{Nataf2013} used a compilation of high-quality HST observations to measure the intrinsic RGBb brightness for 72 Galactic GCs. They explored how the RGBb brightness varied with cluster metallicity, adopting the spectroscopic metallicities of \citet{Carretta2009}, which are the same as those on which our CMD-based approach is calibrated.  As shown in the left panel of Figure~\ref{fig:SC1-rgbb}, they find the luminosity of the RGBb varies by $\sim2$~magnitudes over the interval [Fe/H]$\sim -2.5$ to $-0.5$.

In the right-hand panels of Figure~\ref{fig:SC1-rgbb}, we show the $V$-band RGB luminosity functions for the three GCs, where HB stars have been excluded with a simple cut in colour so that the RGBb feature is more clearly visible. We measure the brightness of the RGBb in each cluster using an automated peak-finding algorithm. Adopting the distance and reddening values measured previously (see Sec.~\ref{method-met}), we calculate the absolute $V$-band\footnote{Standard $UBVRI$ magnitudes are output by Dolphot using the transformations presented by \citet{Sirianni2005}.} magnitudes of these features, which are indicated in the figure by dotted lines. In the left panel, we compare our RGBb and [Fe/H] measurements with the trend defined by MW GCs \citep{Nataf2013}, finding remarkable consistency.  The agreement of this secondary metallicity indicator with our earlier measurements based on RGB curvature lends further support to the reliability of our method. Moreover, because SC1 and SC2 are both low-metallicity clusters for which the red-HB-fitting method was used, this consistency is especially reassuring (see Table~\ref{tab:results} flags).

\section{HB morphology}\label{HBmorph}  

\subsection{HB index}
The HB morphology of the GCs was quantified via the dimensionless HB index defined by \citet{Lee1994} as $(B-R)/(B+V+R)$, where $B$ is the number of blue HB stars, $R$ is the number of red HB stars, and $V$ is the number of stars in the instability strip. A selection box enclosing the HB was defined on the de-reddened CMD by eye. The colour limits of $F606W - F814W = 0.2$ and $0.4$ were chosen to separate the blue, red, and variable stars, following \citetalias{McGill2025}. 
With this definition, a positive HB index represents an extended blue HB, whereas a negative HB index indicates a red HB. 
The measured HB indices\footnote{Note that in the NGC\,6822 cluster SC5 there are a small number of stars that are potentially consistent with an extended blue HB (see Fig.~\ref{fig:CMDs-NGC6822}), however, these were excluded from our selection, as they coincided with the blue plume present in the surrounding field. Including these stars leads to a higher HB index of $-0.3\pm 0.2$ and $\Delta (V-I)= 0.118 ^{+0.019}_{-0.028}$ but overall remains a predominantly red HB morphology. 
}  are indicated in the upper-left corners of Figures~\ref{fig:CMDs-NGC6822} -- \ref{fig:CMDs-NGC185} and are listed in Table~\ref{tab:results}. 
The uncertainties on the HB indices were derived by assuming Poisson statistics. As a result, for GCs with extreme HB index values -- for example, clusters in which all HB stars are red, giving an HB index of $-1.0$ -- the formal uncertainty is zero. 
It is worth noting that potential contamination from the RGB bump in the selection of red HB stars may lead to a slight underestimation of HB index for clusters where this coincides with the HB level, however this effect is expected to be small.

\subsection{Median colour difference between HB and RGB, $\Delta (V-I)$}\label{deltav-i}
\citet{Dotter2010} introduced an alternative metric for quantifying the HB morphology based on the difference between the median colours of HB and RGB stars, denoted as $\Delta(V-I)$. Their analysis of 60 MW GCs using data from the ACS Treasury Survey of Galactic GCs illustrated that $\Delta(V-I)$ correlates well with the HB index while being less susceptible to saturation at extreme HB morphologies. For example, if all stars lie redward of the instability strip, the HB index will be $-1.0$ regardless of how those stars are distributed along the red HB, whereas $\Delta(V-I)$ remains sensitive to that distribution. For this reason, we also computed $\Delta(V-I)$ for our sample as an additional measure of HB morphology.  

To measure $\Delta(V-I)$, it was first necessary to transform the photometry from the native HST filters, $F606W$ and $F814W$, to standard $V$ and $I$ magnitudes. For clusters observed with the ACS, this was straightforward, as standard $UBVRI$ magnitudes are provided by Dolphot using the transformations presented by \citet{Sirianni2005}. However, for the WFC3 observations, such transformations are not available within Dolphot. We therefore applied the empirically-derived transformations of \citet{Harris2018} in these cases. 

Once the photometry had been transformed to $V$ and $I$, a selection box was defined by eye to enclose the HB, following the same approach adopted for the HB index measurements. A second selection box was then defined to isolate the RGB at the level of the HB ($V_{HB}$). The difference between the median colours of these two populations was subsequently measured, and the resulting values are listed in Table~\ref{tab:results}, along with uncertainties estimated via bootstrapping with replacement.  

\begin{figure}
\centering
\includegraphics[width=\columnwidth]{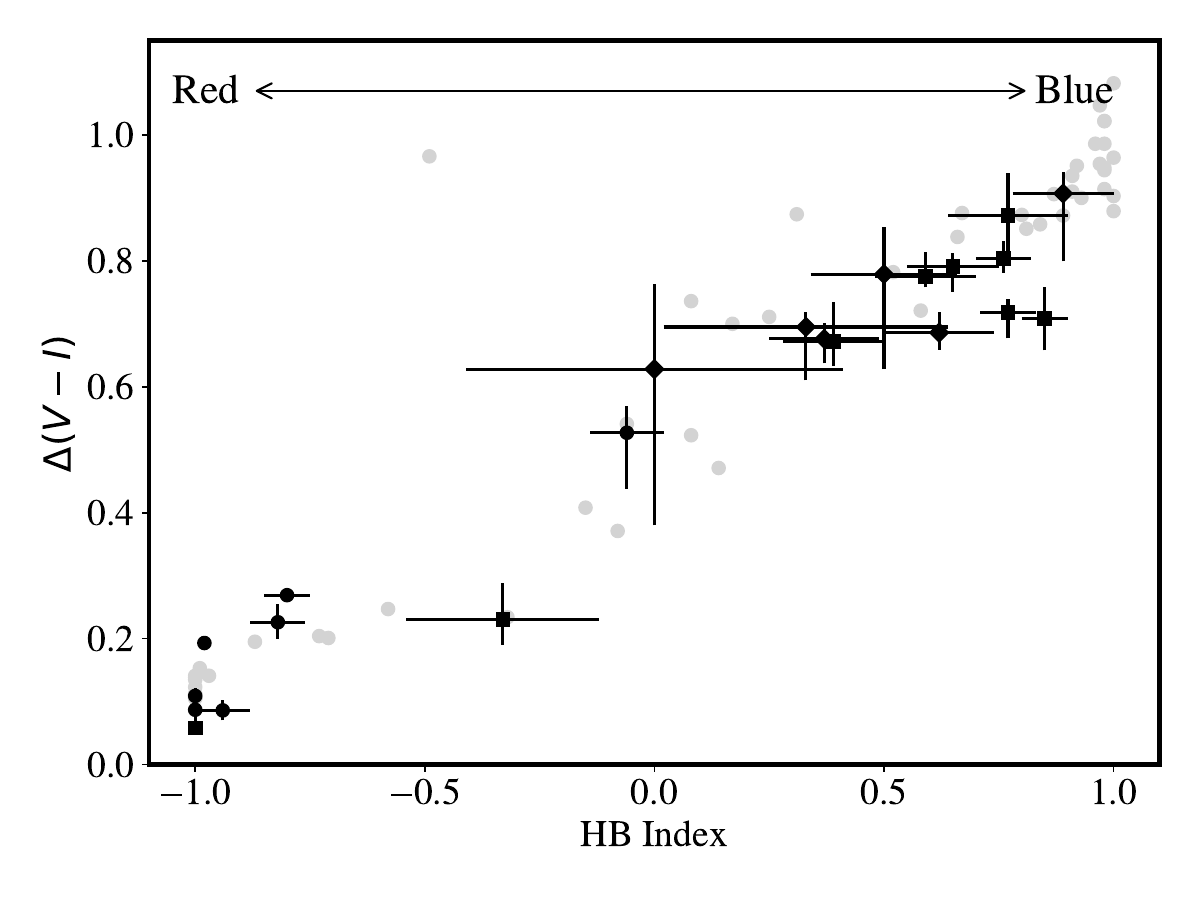}
\caption{HB index vs $\Delta (V-I)$. Measurements from this work are shown in black; MW GCs \citep{Dotter2010} are shown in light grey points.}
\label{fig:HB-morph-methods-comp}
\end{figure}

Figure~\ref{fig:HB-morph-methods-comp} shows the correlation between $\Delta (V-I)$ and HB index, with the MW sample from \citet{Dotter2010} also shown as light grey points. The two indices do indeed correlate well, and the saturation of the HB index becomes evident at extreme red (HB index $\sim -1.0$) and blue (HB index $\sim +1.0$) values. 
Another advantage of using $\Delta (V-I)$ to quantify the HB morphology is that it is unaffected by uncertainties in $E(B-V)$, which influence the placement of the boundaries separating `red', `blue' and `variable' stars for the HB index measurement. 

\section{Sizes and luminosities}\label{sizes} 

For each cluster, the integrated light profile was measured in the F606W band using the \textit{photutils} package \citep{photutils} with concentric apertures of increasing radius.  Except for the notably elongated cluster SC7 (see Section~\ref{SC7}), all clusters appear visually round, and so circular apertures were adopted. Bright contaminants were masked. The cluster centre was determined using the \textit{centroid\_com} function, which calculates the centre-of-mass of the image.  Typical background annuli of width $\sim5$ arcsec were used to determine the mean background flux. The exact width and position of these annuli were selected on a case-by-case basis to avoid bright contaminants. After subtracting the background, the total extent of the cluster was determined by identifying the radius at which the enclosed flux flattened out. 

Measurements of the cluster half-light radii (i.e., the radii enclosing half of the total flux) and \textit{V}-band magnitudes\footnote{As before, magnitudes here were converted from the native HST filters via transformations from \citet{Sirianni2005} and \cite{Harris2018} for the ACS and WFC3 observed clusters, respectively.} are reported in Table~\ref{tab:results}.
Uncertainties in the measured sizes and luminosities of the GCs are expected to be small, with the choice of the radius enclosing the total flux likely contributing the dominant source of error. To estimate this uncertainty, we selected by eye a range of radii over which the enclosed flux could reasonably be said to have plateaued. The typical uncertainties arising from the different choices are estimated to be $\sim 0.05$ mag in
the total apparent magnitude and $\sim 0.4$ arcsec in the angular radius. After also accounting for the uncertainties in distance and reddening,\footnote{For the four GCs (Hubble-VII, Hodge 1, FJJ2 and FJJ3) for which we do not obtain measurements of reddening and distance, we adopt the literature values of the distance of the host galaxy itself. Specifically, we assume a distance of $501$ kpc for NGC\,6822 (see Section,~\ref{6822-distances}) and, for NGC\,147 and NGC\,185, $734.5^{+21}_{-20}$ and $648.6\pm 18$ kpc, respectively \citep{Savino2022}; reddening values are taken from \citet{Schlegel1998}, as recalibrated by \citet{Schlafly2011}.} the typical uncertainties on the absolute magnitudes and physical half-light radii are $\sim0.12$ mag and $\sim0.4$~pc, respectively.

\begin{figure}
\centering
\includegraphics[width=\columnwidth]{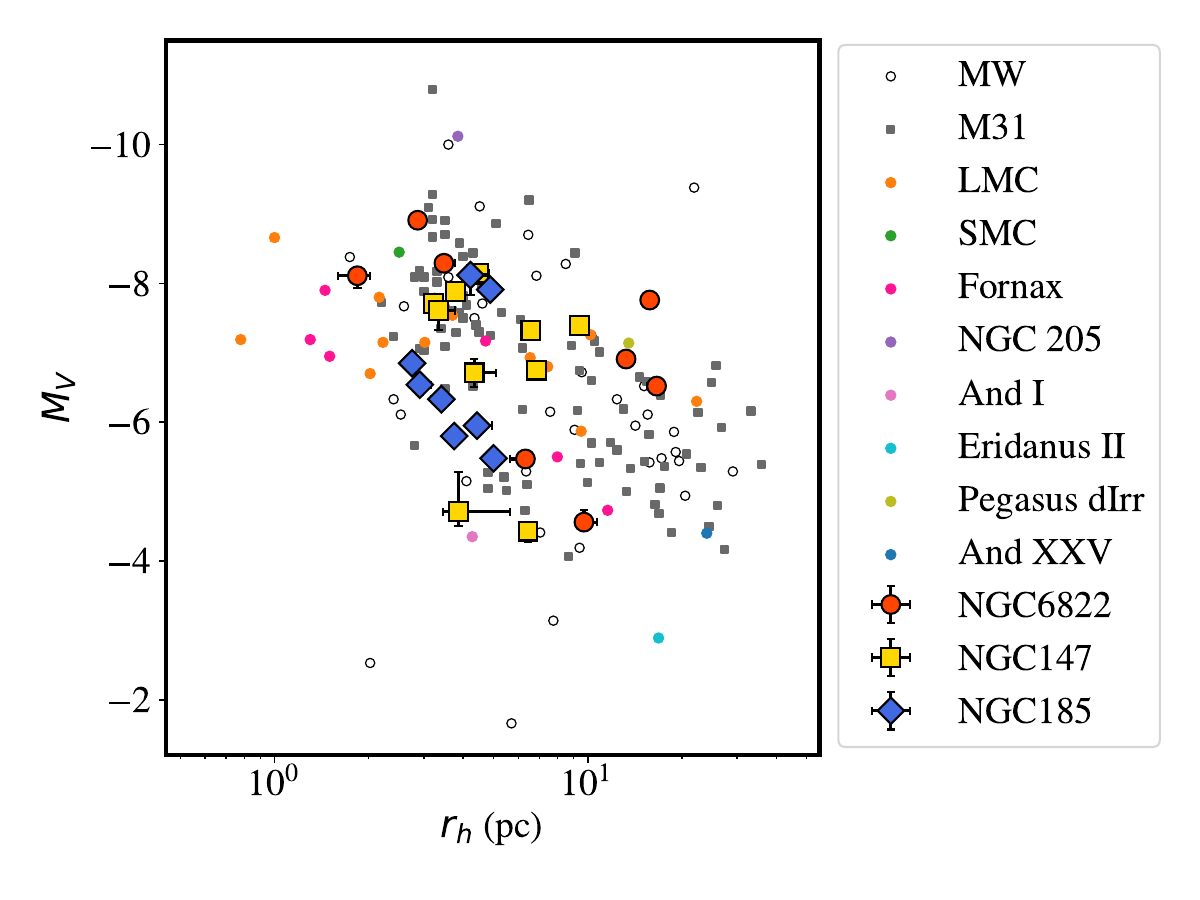}
\caption{
 $M_V$ vs $r_h$ for GCs associated with NGC\,6822, 147 and 185. Outer halo GCs in M31 and MW, as well as those hosted by various LG dwarfs, are also included and listed in the legend. M31 GC data are taken from \citet{Mackey2019mnras}; MW and dwarf GC data are taken from the Local Volume Database \citep[][\url{https://github.com/apace7/local_volume_database}, and citations therein]{Pace2025} with the addition of And~XXV (Crociati et al. in preparation).
}
\label{fig:Mv-rh}
\end{figure}

Figure~\ref{fig:Mv-rh} shows the half-light radius as a function of absolute magnitude for the GCs measured in this study. Outer-halo GCs in M31 and in the MW, as well as clusters hosted by a range of LG dwarf galaxies, are also shown for comparison. 
Overall, the GCs in our sample span a considerable range in magnitude and size, but they generally occupy the same region of parameter space as other LG GCs.  
NGC\,6822 is notable, as it has three extended GCs (SC1, SC2, SC4; $r_h> 10$pc), whereas no such objects are present in NGC\,147 or NGC\,185.  These extended clusters are also the most remote in projection from the centre of NGC\,6822. Most of the other extended clusters shown in Figure~\ref{fig:Mv-rh} reside in the outer halos of the MW and M31.  SC1 is the second most luminous extended GC known, after NGC\,2419, which lies at a galactocentric radius of $\sim95$~kpc in the MW halo. NGC\,2419 is well known for its significant chemical peculiarities \citep[e.g.,][]{2012MNRAS.426.2889M, 2012ApJ...760...86C}, which have led to suggestions that it has been accreted. Other extended GCs in Figure~\ref{fig:Mv-rh} include Reticulum, an outlying GC in the LMC, and the central star clusters of the Eridanus~II and Andromeda~XXV dwarfs. 

The sizes of four GCs in NGC\,6822 (SC1, SC2, SC3, and SC4) were previously measured by \citet{Hwang2011} using a method similar to that adopted here. Comparing angular sizes, we find good agreement between the two studies, with our sizes being slightly larger, on average, by $\sim 0.1$ arcsec.  There are some discrepancies in the total magnitudes measured by \citet{Hwang2011} and those presented here; however, these largely reflect differences in the assumed line-of-sight distances to the GCs. When we adopt the distance and reddening values assumed by \citet{Hwang2011}, we find good agreement, with our measurements being, on average, $\sim 0.11$ mag fainter. 
The total magnitudes of the remaining three NGC\,6822 GCs (SC5, SC6 and SC7) were measured by \citet{Huxor2013}.  Comparing those values with our results, again adopting the distances and reddening assumed in that work, we similarly find our measurements to be slightly fainter,  by $\sim 0.07$ mag.  In addition, the \textit{V}-band magnitudes of all the NGC\,6822 GCs were consistently remeasured by \citet{Veljanoski2015}, whose results also agree well with ours, with a mean offset of only $\sim 0.06$ mag.  

More recent measurements of sizes and luminosities of four NGC\,6822 GCs (Hubble-VII, SC3, SC6, and SC7) were obtained by \citet{Howell2025} using {\it Euclid} ERO data. Their values are in reasonably good agreement with those here, although they again find, on average, slightly larger sizes (by $\sim0.2$ arcsec) and fainter magnitudes (by $\sim0.2$ mag).  

The total magnitudes we measure for the NGC\,147 GCs agree well with those reported by \citet{Veljanoski2013}, with our values being on average $\sim 0.1$ mag fainter. The only exception is Hodge\,4, which we find to be significantly fainter than previously reported, by $\sim1$ mag. 
Hodge\,4 is a very faint, diffuse cluster located in a particularly crowded field, and the discrepancy may reflect differences in our treatment of foreground contaminants and background subtraction relative to the lower-resolution ground-based study of \citet{Veljanoski2013}.
Our NGC\,185 GC magnitudes also appear to be systematically fainter than the values reported by \citet{Veljanoski2013}, by $\sim0.15$ mag on average.

\subsection{SC7}\label{SC7}
\begin{figure}
\centering
\includegraphics[width=0.9\columnwidth]{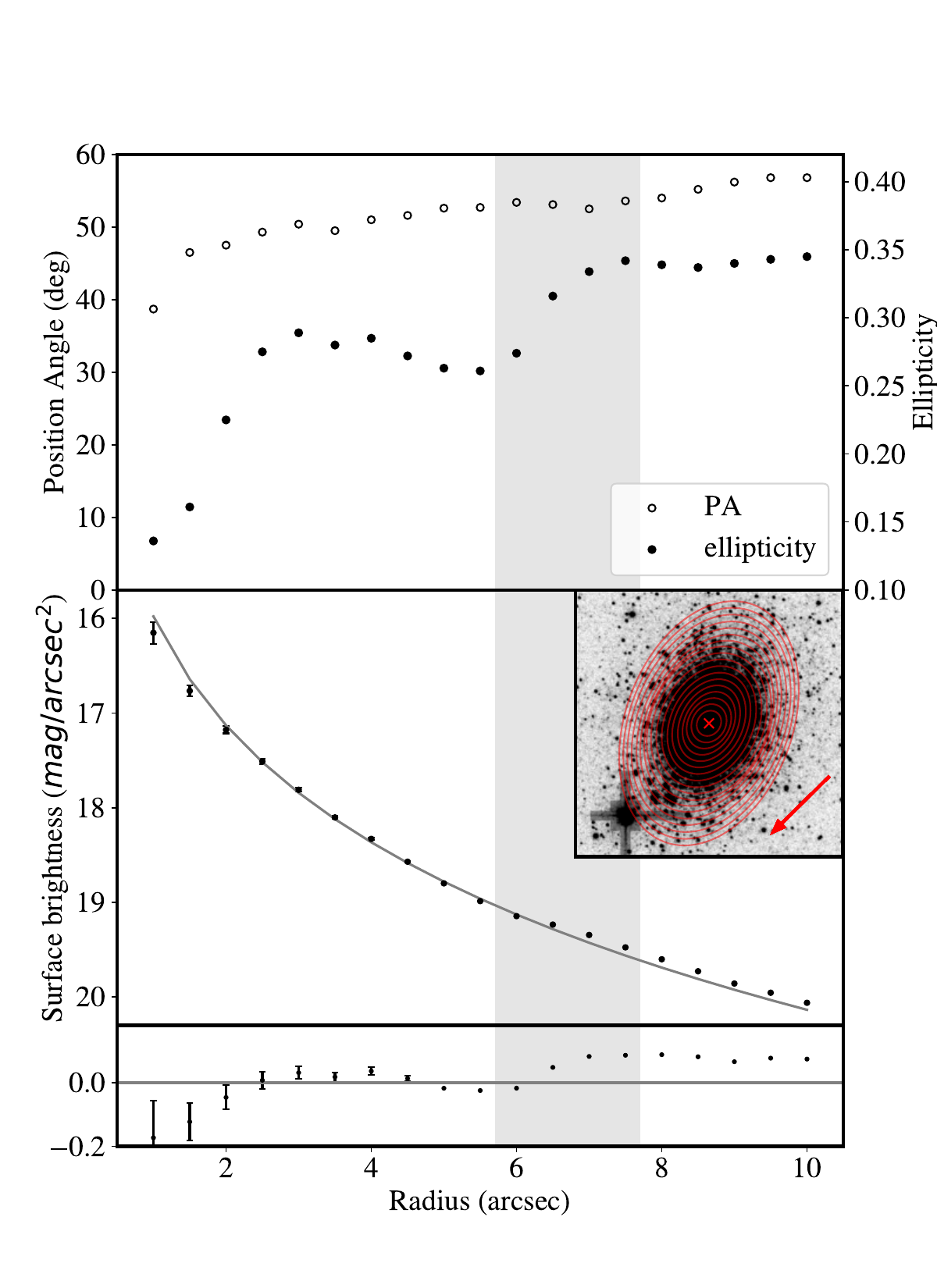}
\caption{
\textit{Top:} Position angle (open circles, left axis) and ellipticity (black circles, right axis) of SC7 as a function of radius. \textit{Middle:} Surface brightness profile of SC7 in F606W. \textit{Bottom:} Residuals from power-law fit to the surface brightness profile. The shaded grey region highlights the increase in ellipticity and the corresponding feature in the surface brightness profile. The inset in the middle panel shows the F606W image of SC7, with the ellipse fits overlaid in red; the arrow points toward the centre of NGC\,6822. 
}
\label{fig:SC7}
\end{figure}

As noted previously (Sec.~\ref {n6822-bkgd}), the NGC\,6822 cluster SC7 is unusually massive, lacks an $\alpha-$enhancement and exhibits a striking elliptical shape. Owing to this morphology, the enclosed light was measured with concentric elliptical apertures, rather than the circular ones adopted for the other GCs in our sample. 
The ellipse parameters were determined from the mean ellipticity and position angle measured over a radial range of $1 - 10$ arcsec using the \textit{ApertureStats} class in \textit{photutils}. 
This routine computes statistics of the pixel distribution within a given aperture, including geometrical parameters such as ellipticity and orientation, determined from a 2D Gaussian with the same second-order moments as the observed light distribution. Bright contaminants were first masked on the F606W image, and the background was measured in an outer sky annulus and subtracted. 
\textit{ApertureStats} was then applied iteratively, starting from an initial aperture defined by eye, until the measured orientation and ellipticity converged to within $0.1^\circ$ and 0.001, respectively. Over the adopted radial range, the mean position angle and ellipticity were found to be $51^\circ$ and 0.29. These values are in very good agreement with previous estimates of $50^\circ$ and 0.25 derived by \citet{Huxor2013} using ground-based data. 

Interestingly, the ellipticity of SC7 begins to rise quite steeply around $\sim 6$ arcsec before flattening again around $\sim 7.5$ arcsec. This increase coincides with an apparent excess of light in the surface brightness profile at a similar radius, which is shown in Figure~\ref{fig:SC7}. 
The top panel shows the ellipticity and position angle as a function of radius, while the bottom panel presents the surface brightness profile measured using concentric elliptical apertures with ellipticities and position angles determined above. To better illustrate the behaviour of the 
surface brightness profile in this radial range, we fit a power law to the profile between $2$ and $6$ arcsec. The residuals with respect to this fit are shown in the bottom panel of Figure~\ref{fig:SC7}. 
The coincident increase in ellipticity and the excess surface brightness are highlighted by the shaded region. Although there are no sharp features in the radial behaviour of the position angle, it does exhibit a subtle increase with radius, together with a slight bump across the radial range of interest.
Taken together, these features provide tantalising evidence that SC7 may be experiencing tidal perturbation. Similar signatures have been identified in the radial profiles of the dE galaxies M32 and NGC\,205, and simulations have shown that such features can arise naturally through the action of tides \citep[e.g.][]{Choi2002}. This interpretation is also consistent with the orientation of the elongation, which points towards the centre of NGC\,6822 (see the red arrow in Fig.~\ref{fig:SC7}).

SC7 has been noted previously for its similarities to $\omega$ Centauri, in terms of its large luminosity, mass and elliptical shape \citep{Howell2025}. It also resembles M19 (NGC\,6273), another luminous Galactic cluster with a highly elliptical morphology \citep[$e=0.27$,][]{Mackey2005}.
Both $\omega$ Centauri and M19 are thought to be the stripped cores of dwarf galaxies accreted onto the MW \citep[e.g.,][]{2003MNRAS346L11B, Pfeffer2021}, suggesting that SC7 could likewise have been accreted onto NGC\,6822. The presence of tidal distortion would suggest that SC7 has been orbiting within the potential well of NGC\,6822 for some time and may therefore date from a merger event that occurred several Gyr ago. This would be consistent with the studies of 
\citet{Zhang2021} and \citet{tantalo2022dwarf-43b}, which find no evidence for substructure in the outskirts of the galaxy that would indicate a very recent merger. Chemo-dynamical studies of SC7 will be required to determine the extent to which rotation also contributes to the flattening of its isophotes \citep{2022MNRAS.512.1584T,2026A&A...708A.286F}, and to search for evidence of a metallicity spread that would strengthen the hypothesis that it is a remnant core of an accreted dwarf galaxy. 

\section{Discussion}\label{discussions}

\subsection{GC Metallicities}\label{gc-met}

Before proceeding to discuss the correlation between HB morphologies and metallicities of the GCs in the three dwarfs, we first comment on the metallicities alone. 

All three dwarfs exhibit similarly broad spreads in GC metallicity.  NGC\,6822 and NGC\,185 contain clusters with metallicities ranging from [Fe/H]$\sim -1.2$ down to the imposed floor at $-2.5$, while NGC\,147 has two slightly more metal-rich clusters, giving an overall range of $-0.7\lesssim $[Fe/H] $\lesssim-2.5$. Additionally, the three dwarfs have very similar mean GC metallicities: $\sim-1.6$ dex in NGC\,6822 and $\sim-1.7$ dex in both NGC\,147 and NGC\,185. These similarities are intriguing given the very different star-formation and chemical-evolution histories of the three dwarf galaxies. 

In particular, NGC\,6822 is a gas-rich dIrr that has experienced continuous star formation to the present day \citep[e.g.,][]{Gallart1996, Fusco2014, nally2024jwst-b12}. The metallicity of its stellar populations varies with age, with typical values of [Fe/H]$\sim -0.4$ to $-0.7$ for stars younger than 100~Myr,  [Fe/H]$\sim -1.3$ to $-1.55$ for stars with ages of $4-8$~Gyr, and [Fe/H]$\sim -1.7$ for stars older than $\gtrsim10$~Gyr \citep{gallart1996local-ea5,2016MNRAS4564315S, tantalo2025dwarf-92e}. Trace populations of stars with metallicities as low as [Fe/H]$\sim -2$ have also been identified \citep{Clementini2003,  2013ApJ779102K,  2016MNRAS4564315S,  ness2025acacias-b36}.  The four innermost GCs (SC3, SC5, SC6 and SC7) have metallicities that overlap with those of the intermediate-age and old field stars, while the three outermost clusters (SC1, SC2 and SC4) are more metal-poor, with [Fe/H]$\lesssim -2$.  Hubble~VII, for which we do not derive a CMD-based metallicity in this paper, has spectroscopic metallicity estimates in the literature ranging from [Fe/H]$=-1.6$ to $-1.95$, indicating that it too is more metal-poor than the bulk of the field stars \citep{cohen1998old-83c,colucci2011detailed-5c8,Larsen2022}. If these metal-poor clusters formed in situ,  this would imply that they formed very early in the evolution of NGC\,6822. However, this interpretation would be in tension with intriguing evidence, from the very red HBs of SC1, SC2 and SC4, that they may be young (see Sec.~\ref{HB-met}).   

The dEs NGC\,147 and 185 formed the bulk of their stellar mass at early times. Using deep HST photometry, \citet{Geha2015} and \citet{Savino2025} determined that $t_{90}$, the look-back time by which 90 per cent of the total star formation had occurred, is $\sim 5$ for NGC\,147 and $8$ Gyrs for NGC\,185.  The presence of metal-rich, red HB GCs in NGC\,147, in contrast to NGC\,185, may suggest that the former experienced a more prolonged period of cluster formation, entirely consistent with their HST-based star-formation histories. 

Various studies have measured metallicities for the field stars in these systems, finding mean values ranging from [Fe/H]$\sim -0.51$ to $-1.1$ for NGC\,147, and from [Fe/H]$\sim -0.98$ to $-1.5$ for NGC\,185 \citep{2010ApJ711361G,Crnojevic2014,2015ApJ79877H}. While NGC\,147 does not exhibit a radial metallicity gradient, NGC\,185 does, with stars at larger radii being a few tenths of a dex more metal-poor than those in the inner regions  \citep{Crnojevic2014,2015ApJ79877H}. In both systems, the GCs overlap the metallicity range of the field stars, but extend to much lower values, reaching [Fe/H]$\lesssim -2$.  In contrast to the field stars, there is evidence for a metallicity gradient in the GC system of NGC\,147 but not in that of NGC\,185. In NGC\,147, the innermost four clusters (Hodge\,2, Hodge\,3, Hodge\,4, SD10) have a mean metallicity of $\sim - 1.2$ dex, compared with $-2.3$ for the outermost five clusters (SD5, SD7, PA1, PA2, PA3).  In NGC\,185, the innermost three GCs (FJJ1, FJJ4 and FJJ5) have a slightly higher mean metallicity ($\sim -1.5$ dex) than the outer clusters ($\sim 1.9$ dex), but the dispersion in both groups is significant.

Theoretical models predict that both the mean metallicity and the metallicity dispersion of a GC system scale weakly with the host halo mass \citep[e.g.][]{Choksi2018}. The similar mean metallicities of the GCs in the three dwarf galaxies in our sample could suggest that they inhabit similarly massive halos. On the other hand, as shown in Fig.~2 of \citet{Choksi2018}, their models predict a metallicity dispersion of $\sim 0.4$ dex for $M_\star \sim 10^9 M_\odot$, whereas we observe substantially larger dispersions of $\sim 1$ dex in all three dwarfs. Although \citet{Choksi2018} do note that their models systematically underestimate the metallicity dispersions by $\sim 0.1 - 0.2$ dex relative to observations, the discrepancy found here is considerably larger. 

\subsection{HB morphology and Metallicity}\label{HB-met}

\begin{figure*}
\centering
\includegraphics[width=\textwidth]{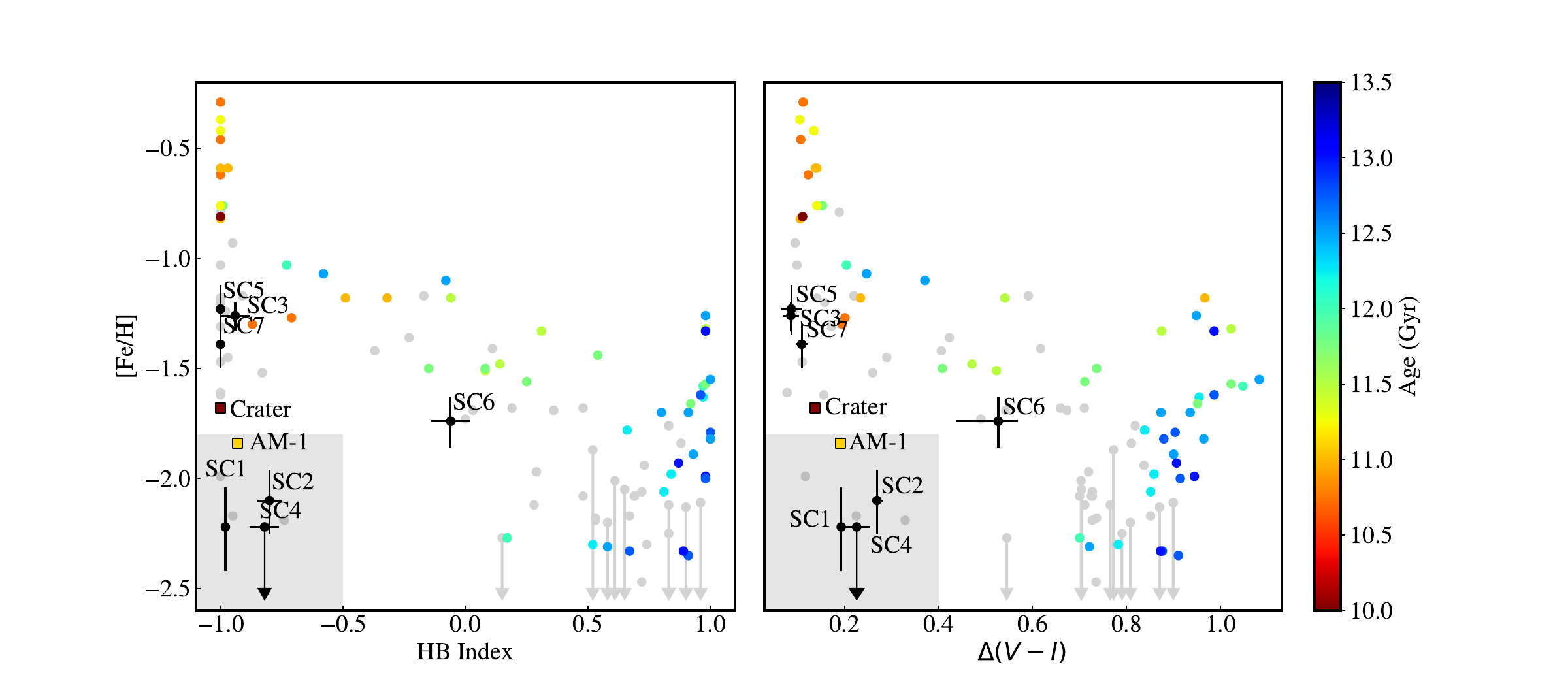}
\caption{Metallicity vs HB morphology. NGC6822 GCs are shown as labelled black points. M31 outer halo GCs are plotted as light grey circles. MW GCs are presented as circles coloured by cluster age, AM-1 and Crater are shown as square symbols. The shaded region highlights anomalous GCs with very low metallicities ([Fe/H] $\lesssim -1.8$) and exclusively red HBs (HB index~$ <-0.5$; $\Delta (V-I) < 0.4 $). Left: HB index; right: $\Delta (V-I)$.}
\label{fig:HB-vs-met-6822}
\end{figure*}

In their seminal study of the MW GC population, \cite{Searle1978} used the spread in HB morphologies at a given metallicity to argue for a prolonged period of accretion in the 
build-up of the Galactic outer halo. This picture has been reinforced by many subsequent studies \citep[e.g.,][]{Zinn1993, Mackey2005}, with strong evidence put forth for red-HB outer-halo GCs having somewhat younger ages \citep[e.g.,][]{Stetson1999, Dotter2010, Dotter2011, Weisz2016}. In M31, direct evidence for an association between red-HB outer-halo GCs and recent accretion was presented by \citetalias{McGill2025}, who showed that, in their sample of M31 halo GCs, almost all clusters with red HBs can be linked, at least tentatively, to tidal debris features.  Here we examine the behaviour of HB morphology and metallicity in the GCs of our dwarf galaxy sample, and explore how they compare to the suspected accreted GCs in the MW and M31 halos.

\subsubsection{NGC~6822} \label{6882-Hb-met}

Figure~\ref{fig:HB-vs-met-6822} shows the HB-morphology-metallicity diagrams for NGC\,6822, using the two different measures of HB morphology.  
For comparison, outer-halo GCs in M31 from \citetalias{McGill2025}
are also shown in light grey, along with a sample of MW GCs coloured according to their main-sequence turn-off (MSTO) ages \citep{VDB2013}. Metallicities for the MW clusters are taken from the spectroscopic measurements by \citet{Carretta2009}. These metallicity measurements were also used to calibrate the empirical relationships derived by Mackey et al. (in prep), so the MW, M31, and dwarf galaxy GC metallicities are all on the same scale and can be directly compared. HB indices for MW GCs are taken from \cite{Mackey2005}, and $\Delta (V-I)$ values from \cite{Dotter2010}.

As discussed in Sec.~\ref{gc-met}, the GCs associated with NGC\,6822 are uniformly metal-poor, ranging from [Fe/H]$\sim -1.2$ down to the imposed floor of $-2.5$ dex. Furthermore, almost all of them have very red HB morphologies (HB index$<-0.5$), with the sole exception of SC6, which has an intermediate HB morphology (HB index = $-0.06$).  Even SC6, however, appears to have a slightly lower metallicity than MW GCs with comparable HB morphology.  Notably, the three clusters (SC1, SC2 and SC4) with very low metallicities ([Fe/H]~$<-1.8$) have very red HBs, much redder than any of their similarly metal-poor MW halo counterparts, aside from AM-1.
These GCs are highlighted by the shaded region of the plot. 
If age is the second parameter, in addition to metallicity, that governs HB morphology \citep[e.g.][]{Dotter2010,Dotter2011}, then this would suggest that these clusters are likely to be younger than those with bluer HBs at similar metallicities.
The only known GC in the MW with such a red-HB at these low metallicities is AM-1. First identified as a GC by \citet{1979ApJ...227L.103M}, AM-1 is one of the most distant known Galactic GCs \citep[$\sim120$ kpc, e.g.][]{2021MNRAS.505.5957B}, and along with one or two other clusters on similarly high-energy retrograde orbits, has been suggested to be associated with the Elqui stream \citep{Massari2019,2026A&A...706A.130D}. 
A MSTO age determination suggests that AM-1 is slightly younger \citep[][$\sim11$ Gyr]{2008AJ....136.1407D} than typical outer-halo GCs in the MW. 
In addition to AM-1, Crater also exhibits similar HB morphology \citep[HB index $= -1.0$ and $\lbrack \text{Fe/H} \rbrack = -1.68$,][]{Kirby2015}. While it is more metal-rich than any of the metal-poor red-HB GCs found in NGC\,6822, it is still $\gtrsim 0.5$ dex more metal-poor than almost all other MW GCs with such red HB morphologies. As Crater was not included in the sample of \citet{Dotter2010}, we measure its $\Delta (V-I)$ value here  
using data from the HST program GO-13746 (PI: Walker), comprising F606W and F814W imaging that reaches to the MSTO. We follow the photometric procedure and methodology outlined in Sections~\ref{obs}~and~\ref{HBmorph}, and find $D(V-I) = 0.14 \pm 0.01$. 
Like AM-1, Crater lies at a very large Galactocentric distance in the outer MW halo ($\sim145$ kpc) and has also been shown to be rather young ($\sim7.5$ Gyr), with a suspected accretion origin \cite{Weisz2016}.  
The similarity in HB morphology to MW GCs with well-determined younger ages reinforces our suggestion that SC1, SC2 and SC4 are likely to be younger than typical Galactic GCs.   
Other examples of GCs with anomalously red HBs and low metallicity were recently identified in the outer halo of M31 by \citetalias{McGill2025}, all of which are at least tentatively associated with substructure. The strong similarities between SC1, SC2 and SC4 in NGC\,6822 and those associated with tidal streams in M31 could indicate either that M31 accreted an NGC\,6822-like dwarf in its recent past, or that both M31 and NGC\,6822 have more recently accreted similar low-mass systems. Given the remote outer-halo locations of these clusters in NGC\,6822, the latter scenario may be more likely. 

It is worth briefly considering the possibility that the metallicities of these metal-poor red-HB clusters may be underestimated due to unknown systematics since no comparably metal-poor red GCs are observed in the MW sample (Mackey et al. in prep.), or if, for example, their $\alpha$-abundances differ from the typical values seen for MW GCs (as discussed in Section~\ref{results}). However, even if they are underestimated by as much as $\sim 0.3$ dex, as might be expected for solar-scaled $\alpha$-abundances, their HB morphologies would still be remarkably red for their metallicities. 
In addition, the brightness of the RGBb (see Sec.~\ref{RGBbump}) provides independent support for our inference that both SC1 and SC2 are very metal-poor.

\begin{figure*}
\centering
\includegraphics[width=\textwidth]{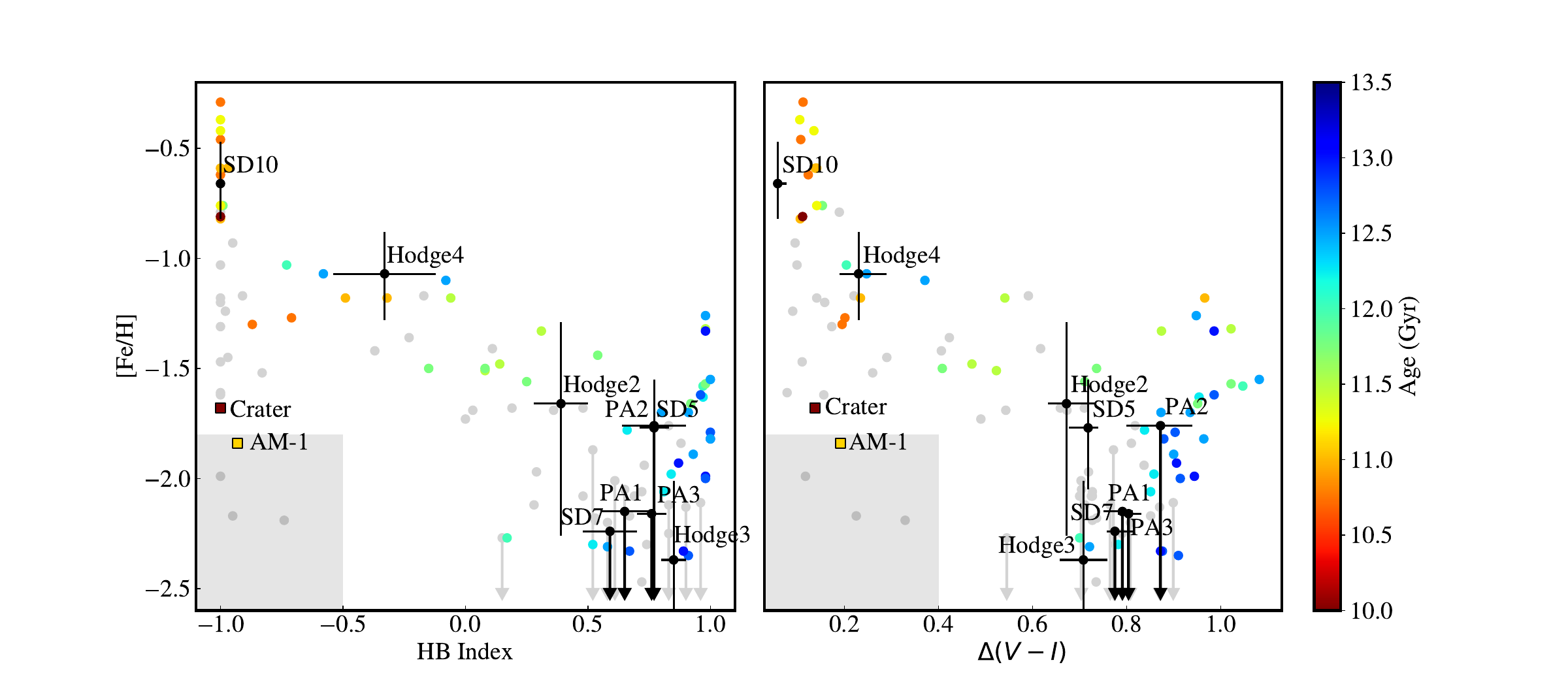}
\caption{Metallicity vs HB morphology for NGC\,147 GCs, as in Figure~\ref{fig:HB-vs-met-6822}.}
\label{fig:HB-vs-met-147}
\end{figure*}
\begin{figure*}
\centering
\includegraphics[width=\textwidth]{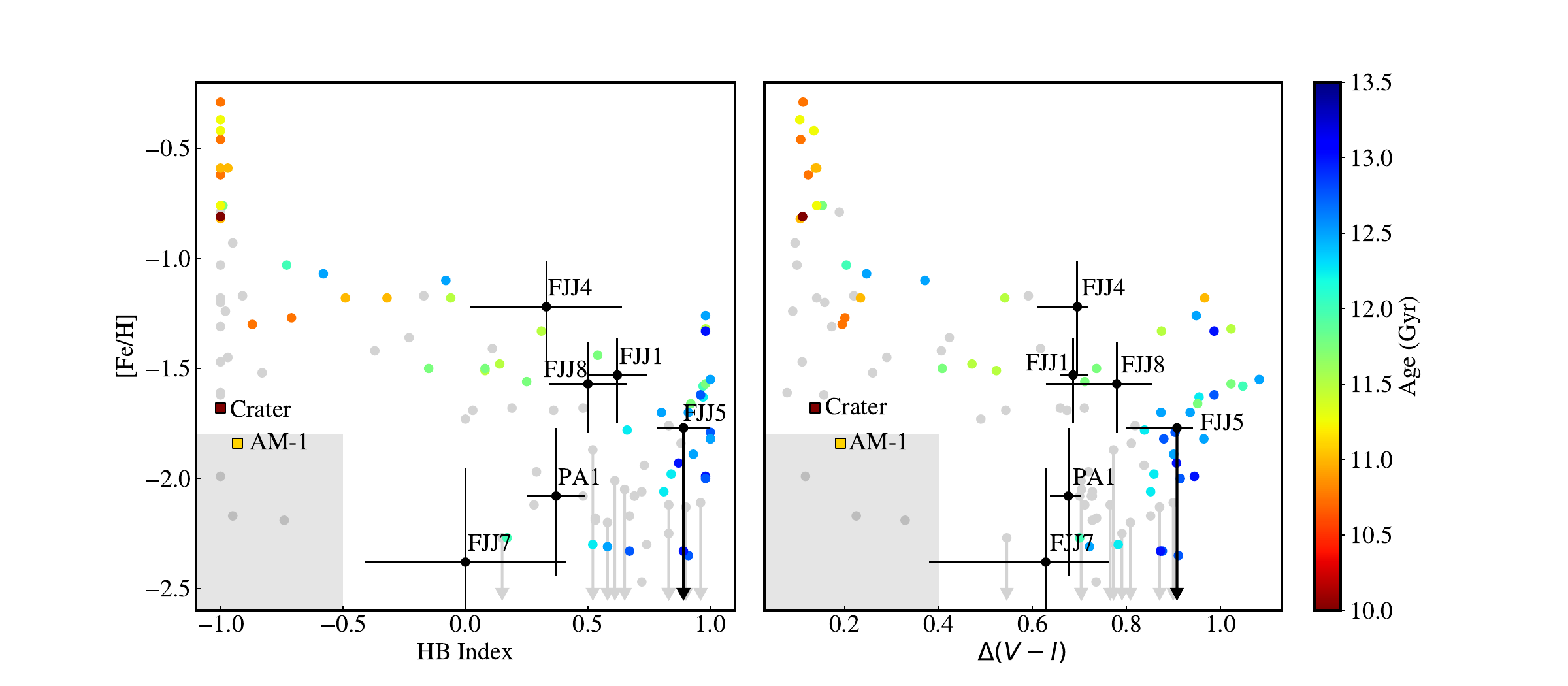}
\caption{Metallicity vs HB morphology for NGC\,185 GCs, as in Figure~\ref{fig:HB-vs-met-6822}.}
\label{fig:HB-vs-met-185}
\end{figure*}

Given the complexity of HB morphology, which is expected to depend not only on 
metallicity and age, but also on additional parameters such as helium (He) abundance and RGB mass loss \citep[e.g.,][]{Gratton2010}, previous studies have adopted more sophisticated methods to quantify it. In particular, \citet{Milone2014} defined the parameters $L_1$ and $L_2$ as the difference between the 96th percentile of the HB colour distribution and the median RGB colour, and the difference between the 4th and 96th percentiles of the HB colour distribution, respectively. They showed that the colour of the reddest HB stars ($L_1$) is more sensitive to metallicity and age, whereas the spread in colour ($L_2$) depends more strongly on cluster luminosity and internal He-abundance spread. In this work, we quantify HB morphology using the more traditional HB index and $\Delta (V-I)$, as these metrics are more robust for our extragalactic dataset, for which the photometric precision is lower than that available for MW GCs. Furthermore, \citet{Milone2014} showed that $L_1$ correlates reasonably well with both HB index and $\Delta (V-I)$, suggesting that these measures are generally sufficient for our analysis. 
However, for the three GCs in NGC\,6822 with anomalously red HBs and low metallicities, we also consider how adopting this more detailed parametrisation of HB morphology would affect our results. We find that the $L_1$ values for SC1 ($0.10\pm 0.01$), SC2 ($0.13 \pm 0.02$) and SC4 ($0.14 \pm 0.03$) are indeed lower (indicating redder HBs) than those of similarly metal-poor ([Fe/H]~$\lesssim -2.0$) MW GCs. In the \citet{Milone2014} sample, the most metal-poor cluster with a similarly low value of $L_1$ is AM-1 ($0.093\pm 0.07$). We additionally find that Crater also exhibits a low value of $L_1 = 0.09 \pm 0.02$, thus further strengthening the idea that these clusters may be younger than those typically found in the MW. 

Age estimates for four clusters in NGC\,6822 have previously been made using Lick indices, including all three low-metallicity red-HB GCs \citep{Hwang2014}. Their results suggest that SC1 and SC2 are as old as typical Galactic GCs ($\sim 12$ Gyrs), while SC4 (and also SC3) are estimated to be slightly younger. However, these measurements carry significant uncertainty ($\sim 2-3$ Gyrs), as does the Lick-index method of age determination more generally \citep[e.g.,][]{2008ApJS..177..446G}.  Robust age estimates for these very metal-poor red-HB GCs will therefore require deep MSTO photometry. Importantly, their ages will place a limit on the timing of any putative merger event, since such a merger could not have occurred before the clusters formed within their original host.

\subsubsection{NGC~147 and NGC~185}

The HB-morphology-metallicity diagrams for the GCs of NGC\,147 are shown in Figure~\ref{fig:HB-vs-met-147}. The clusters in this dwarf galaxy span a wide range of metallicities ($-0.7< $ [Fe/H] $<-2.5$) and HB morphologies, including both extremely blue and extremely red HB clusters. That said, the majority of GCs are metal-poor and have blue HB morphologies. 
Overall, this trend closely follows that seen in the MW halo sample, and no anomalously metal-poor red-HB GCs, such as those found in the outer halo of M31 and in NGC\,6822, are present.
Age estimates for three of the NGC\,147 GCs (Hodge 2, SD5 and SD7) were derived by \citet{Sharina2009} using Lick indices. Their results suggest that SD7 may be younger ($8\pm 2$~Gyr) than Galactic GCs at similar metallicity, but this is at odds with its reasonably blue HB morphology, which instead suggests an old age.  It should be noted, however, that the age estimates of \citet{Sharina2009} are highly uncertain; in particular, they emphasise that these clusters may appear young due to artificial filling-in of the Balmer lines. As noted above, precise age estimates from the MSTO would therefore be highly informative.

The GCs in NGC\,185 also generally follow the relationship between HB morphology and metallicity that is observed for MW GCs. One possible exception, indicated in Figure~\ref{fig:HB-vs-met-185}, is FJJ7, which may have a redder HB morphology than MW GCs at similar metallicity. However, as shown in Fig.~\ref{fig:CMDs-NGC185}, FJJ7 has a very sparse HB, making its characterisation highly uncertain. 
On the other hand, \citet{Sharina2009} estimate ages for eight clusters in NGC\,185 and finds evidence that FJJ7 may be significantly younger ($5\pm 2$~Gyrs) than the other clusters in this dwarf, which would be consistent with its redder HB morphology. However, it is also worth noting that the metallicity they derived for FJJ7 ($-0.8\pm 0.2$ dex) is significantly higher than the value determined here. Overall, we find all of the GCs in NGC\,185 to be metal-poor, with metallicities ranging from $\sim -1.2$ dex down to the imposed metallicity floor of $-2.5$ dex, and all exhibit blue HB morphologies (HB index $\gtrsim 0$). 

We previously remarked on the fact that the dwarf galaxies have similar GC mean metallicities and metallicity dispersions despite their very different evolutionary histories. This picture changes when HB morphology is considered alongside metallicity. Although NGC\,147 hosts a slightly higher fraction of red-HB GCs than the neighbouring NGC\,185, its GC system still has considerably bluer HBs on average than that of NGC\,6822. If all of the GCs in NGC\,6822 formed in situ, then its higher fraction of red-HB clusters relative to the two quenched dE satellites is consistent with its more extended star-formation history. Conversely, if the metal-poor red-HB clusters in NGC\,6822 have been accreted, this would imply that it has experienced a more prolonged accretion history than either of the two dEs. 

\subsection{Distances and 3D spatial distribution of GCs}\label{distances}
\begin{figure*}
\centering
\includegraphics[width=\textwidth]{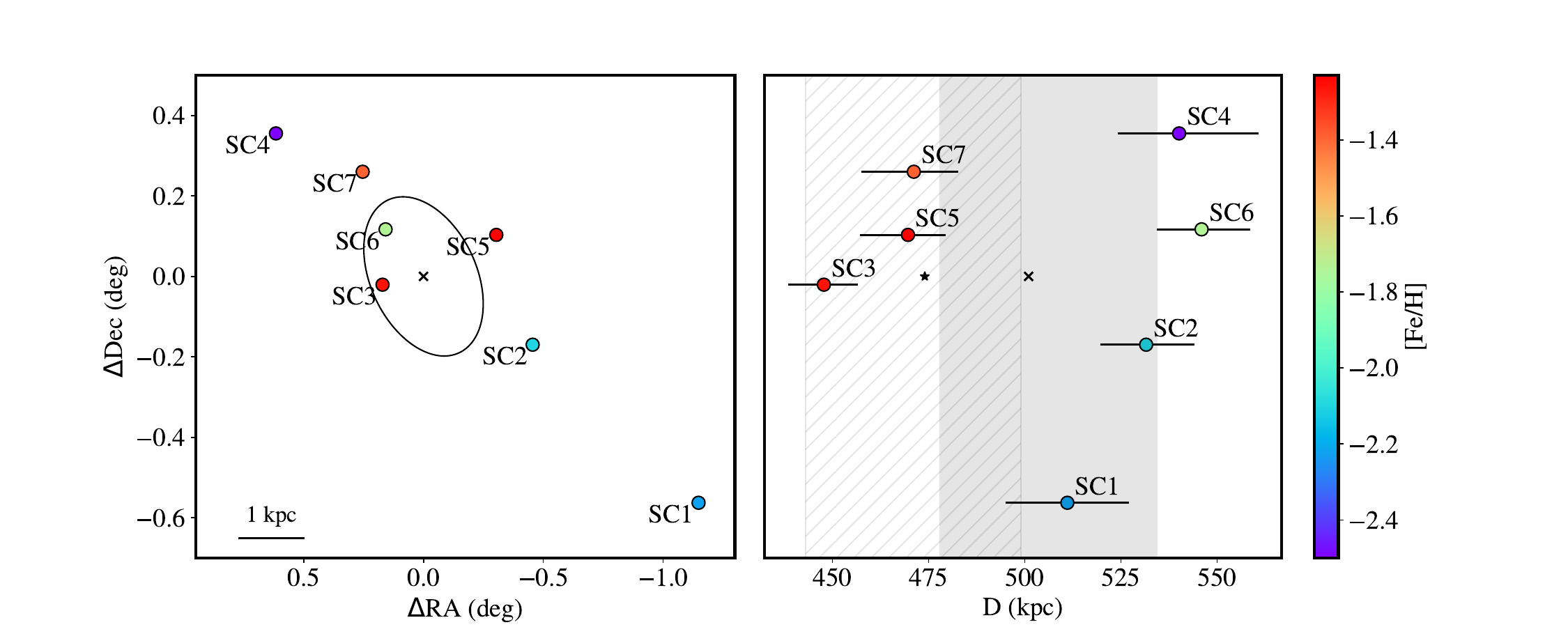}
\caption{Spatial distribution of GCs in NGC\,6822, colour-coded by metallicity. \textit{Left}: projected positions of GCs around NGC\,6822, with the galaxy centre marked by a cross. The half-light ellipse is also shown, as in Fig.~\ref{fig:images}. \textit{Right}: distribution of GCs along the line-of-sight. The solid grey shaded region indicates the 1-$\sigma$ range of I-band TRGB distances reported in the literature, compiled from the NASA/IPAC Extragalactic Database, with the addition of \citet{Higgs2021}. The cross marks the median distance. The hatched region indicates the 1-$\sigma$ range of literature distances measured using Cepheid variable stars, with the star indicating the median.}
\label{fig:dist-6822}
\end{figure*}

\begin{figure*}
\centering
\includegraphics[width=\textwidth]{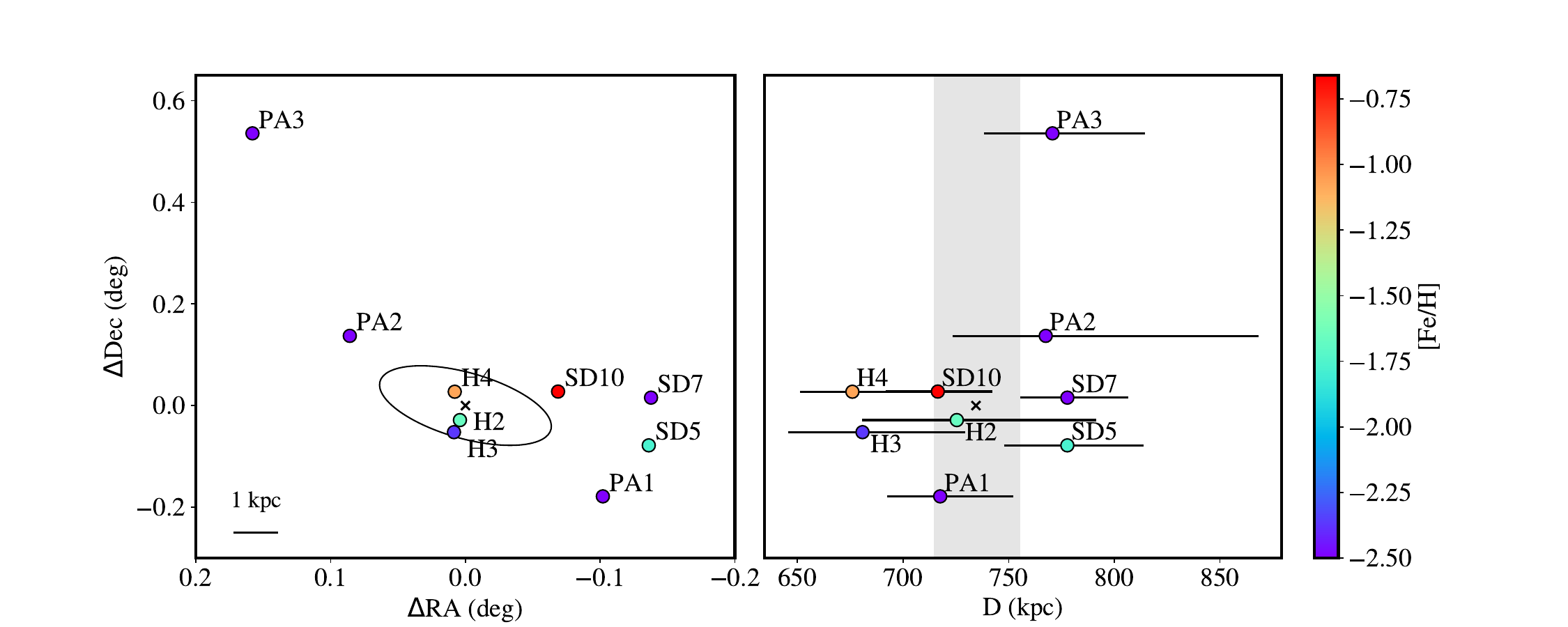}
\caption{Spatial distribution of GCs in NGC\,147, colour-coded by metallicity. \textit{Left:} projected positions of GCs around NGC\,147, with the galaxy centre marked by a cross. The half-light ellipse is also included as in Fig.~\ref{fig:images}. \textit{Right:} distribution of GCs along the line-of-sight. The cross indicates the literature distance to NGC\,147, and the shaded region shows the associated uncertainty \citep{Savino2022}.}
\label{fig:dist-147}
\end{figure*}

\begin{figure*}
\centering
\includegraphics[width=\textwidth]{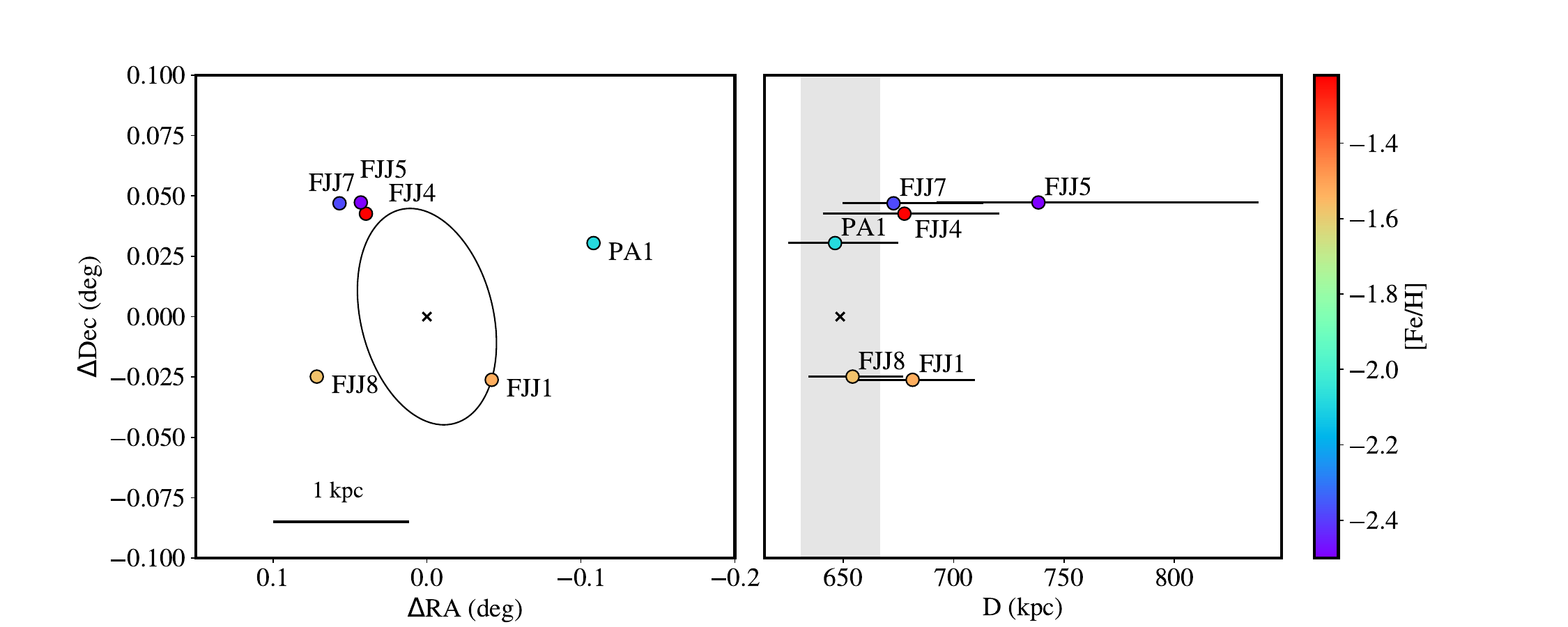}
\caption{Spatial distribution of GCs in NGC\,185, as in Figure~\ref{fig:dist-147}.}
\label{fig:dist-185}
\end{figure*}

\subsubsection{NGC~6822}\label{6822-distances}

Figure~\ref{fig:dist-6822} shows the spatial distribution of the NGC\,6822 GCs, colour-coded by metallicity, both in the plane of the sky (left) and along the line-of-sight (right), using the reddening and distance measurements derived as described in Sec.~\ref{method-met}. 
Although a clear metallicity gradient appears to be present when considering projected positions, the picture becomes more complicated once line-of-sight distances are taken into account.

In the left panel, the centre of the galaxy is indicated by a cross, whereas in the right panel we show the median of the tip of the RGB (TRGB) $I$-band measurements reported in the literature, compiled from the NASA/IPAC Extragalactic Database and supplemented by the more recent \citet{Higgs2021} measurement. We prefer to show the median ($501$~kpc), as opposed to the mean, due to the significant spread in the literature measurements of NGC\,6822's distance: considering $I$-band TRGB measurements alone, the distance measurements span a 1-$\sigma$ range of $\sim60$kpc 
($478-535$~kpc), as indicated by the solid grey shaded region in the plot. 
Discrepancies in literature distance estimates remain just as severe when different methodologies are considered. \citet{Feast2012} find that previous measurements based on the Cepheid period-luminosity relation differ by up to nine per cent. The median Cepheid distance from NED (indicated by a star in the right panel of the figure) is 474 kpc, around 30~kpc closer than the median TRGB distance, and this distance is in reasonably good agreement with the more metal-rich GCs in our sample. 
The 1-$\sigma$ range of Cepheid distance estimates is shown by the hatched region in Figure~\ref{fig:dist-6822}.
To the best of our knowledge, only one RR Lyrae distance measurement is available, from \citet{Clementini2003}, who found a distance of $470 \pm 38$ kpc, in good agreement with average Cepheid estimates.

From Figure~\ref{fig:dist-6822}, it is immediately clear that the GC distances are strongly bimodal, with the metal-poor GCs located at a mean line-of-sight distance of $536$~kpc and the more metal-rich GCs at a mean distance of $463$~kpc. Given the poorly constrained distance to NGC\,6822, either group
could in principle be consistent with the galaxy, and the interpretation of this large offset is therefore not straightforward. We first consider how likely it is that the distances to the metal-poor, red-HB clusters (SC1, SC2 and SC4) could be affected by unknown systematics. As noted previously, there are no GCs in the MW calibration sample with this particular combination of properties, so we cannot rule out the possibility that the distances to the clusters are overestimated. However, theoretical models indicate that the HB of a younger cluster is both redder and brighter than that of an older cluster. Therefore, if age differences are responsible for the red-HB morphologies at low metallicity, the distances to these clusters would be expected to be underestimated rather than overestimated. This would only increase the observed line-of-sight offset between these GCs and the more metal-rich population. Using BaSTI isochrones \citep{Pietrinferni2021}, we calculate that the HB magnitude of a 12~Gyr low metallicity ([Fe/H] $= -2.2$) cluster is $\sim 0.1$ mag fainter than that of an 8~Gyr old cluster. At the median TRGB distance of NGC\,6822 from the literature, this corresponds to a distance underestimation of around $\sim 20$kpc for the younger GC. Furthermore, SC6, which has an intermediate HB and a modest metallicity of [Fe/H]$\sim -1.7$, lies at an even greater line-of-sight distance than any of the red-HB metal-poor clusters. This suggests that the larger distances to the low-metallicity group cannot be wholly explained by systematics associated with the HB morphology. 
   
Another possibility is that the spread in the distances to both the galaxy and its GC population is at least partly real, indicating a genuine line-of-sight depth in NGC\,6822. It has long been known that the Small Magellanic Cloud (SMC) is significantly extended along the line-of-sight despite its small angular size on the sky \citep{Mathewson1986}. For example, \citet{Ripepi2017} measured a depth of $\sim30$~kpc and argue that this is likely the result of tidal interactions with the LMC. A similar line-of-sight depth in NGC\,6822 could account somewhat for the wide range of distances reported in the literature. However, for this isolated dwarf, which shows no recent signs of interactions, the origin of such line-of-sight elongation would be more challenging to explain than in the case of the SMC.  Further investigation will certainly be required to confirm whether it is physical in nature.

\subsubsection{NGC\,147 and NGC\,185}\label{147-185-dist}
Figure~\ref{fig:dist-147} shows the spatial distribution of the NGC\,147 GCs. 
In the right panel, the distance to the galaxy is taken to be $734.5^{+21}_{-20}$~kpc from the recent RR Lyrae study of \citet{Savino2022}.
The mean distance to the GCs is $734.2$~kpc, which agrees remarkably well with the distance of the galaxy itself, suggesting that the clusters are distributed reasonably symmetrically along the line-of-sight. However, there is a significant spread of $\sim 100$ kpc along the line-of-sight, which is likely driven by, or at least affected by, the large uncertainties in the individual GC distance determinations. Indeed, the typical uncertainties are $\sim 40$kpc, almost twice the mean uncertainty for the NGC\,6822 GCs.

It is worth noting that Mackey et al. (in prep) caution that the intrinsic HB luminosities of the most metal-rich clusters ([Fe/H] $\gtrsim -1.0$) deviate from the derived linear relation with metallicity. As a result, applying this method to determine the distances to the most metal-rich GCs may result in a slight underestimation of their distances. 
SD10 is the only GC in our sample within this metallicity regime, and we therefore flag its distance measurement as uncertain. That said, the distance we derive for SD10 is consistent with that of NGC\,147 itself, suggesting that any such underestimation is likely to be small.

The spatial distribution of the GCs in NGC\,185 is shown in Figure~\ref{fig:dist-185}, where the distance to NGC\,185 is taken to be $648.6\pm 18$ kpc, again from the recent RR Lyrae study of \citet{Savino2022}. The GCs appear to lie mostly on the far side of the galaxy, with a mean distance of $670$~kpc. 
This is skewed slightly higher by FJJ5, which has a highly uncertain distance measurement of $739^{+100}_{-46}$~kpc. Excluding this cluster, we find a mean distance of $666$~kpc, which is consistent with the literature distance to NGC\,185 within 1-$\sigma$. As with NGC\,147, a significant dispersion is seen along the line-of-sight; however, again, this is largely due to the significant uncertainties associated with the individual GC distance measurements.

\section{Conclusions}\label{conclusions}

In this paper, we have presented deep HST ACS and WFC3 photometry of the entire GC populations of three Local Group dwarfs, comprising 26 GCs in total: the isolated dIrr galaxy, NGC\,6822, and the M31 dE satellites, NGC\,147 and NGC\,185.  With this data, we have analysed, for the first time, deep CMDs extending below the HB for GCs in dwarf galaxies beyond the immediate surroundings of the MW. From these CMDs, we quantified the RGB and HB morphologies and used empirical relationships to derive metallicities, line-of-sight extinctions and distances. 

Despite their very different evolutionary histories, we find no evidence for strong differences in the metallicity distributions of the three dwarf galaxies,  either in their dispersion or in their mean values. NGC\,6822 and NGC\,185 contain clusters with metallicities ranging from [Fe/H]$\sim -1.2$ down to the imposed metallicity floor at $-2.5$ dex, while NGC\,147 has two slightly more metal-rich clusters, giving an overall range of $-0.7\lesssim $[Fe/H] $\lesssim-2.5$). Their mean metallicity values are likewise very similar, at $\sim-1.6$ dex in NGC\,6822 and $\sim-1.7$ dex in both NGC\,147 and NGC\,185. 
We have extensively compared our CMD-based metallicity and reddening measurements with values reported in the literature, most of which are derived using completely independent techniques. Overall, we find very good agreement, and where discrepancies do arise, they can generally be understood.  

In contrast to their similar metallicities, we find intriguing differences in the HB morphologies of the GCs in the three dwarf galaxies. All but one of NGC\,6822's GCs exhibit red HB morphologies, whereas those in the two dEs are predominantly blue. Remarkably, the three most outlying GCs in NGC\,6822 (SC1, SC2 and SC4) have very low metallicities ([Fe/H]~$<-1.8$) yet very red HBs, a combination not seen in the other dwarf galaxies or in the MW halo GC population. The closest MW analogues are the two outer-halo GCs AM-1 and Crater; however, both of these clusters are more metal-rich for their HB morphology compared to SC1, SC2 and SC4.  While it is possible that the metallicities of these clusters may be affected by systematics, we note that for SC1 and SC2 the values derived from the curvature of the RGB are also supported by empirical relationships between RGB bump magnitude and metallicity \citep{Nataf2013}.  Assuming that age is the second parameter governing HB morphology \citet{Dotter2010,Dotter2011}, we suggest that these GCs are likely younger than their similarly metal-poor MW counterparts, although precise MSTO ages are required to confirm this.  If these clusters are indeed young, their origin is difficult to explain given their very remote locations within NGC\,6822 and their low metallicities, which are significantly below those of the old and intermediate-age stars in NGC\,6822.  We note that these three clusters are strikingly similar to several GCs linked to substructure in the outer halo of M31 \citepalias{McGill2025}, suggesting either the recent accretion of an NGC\,6822-like dwarf by M31, or that both M31 and NGC\,6822 have recently accreted a similar low-mass system. 

While the dEs possess clusters with predominantly blue HBs, NGC\,147 contains two clusters that are more metal-rich, with redder HBs, than any of those found in NGC\,185.  The presence of these GCs may suggest that NGC\,147 experienced a more prolonged period of cluster formation, entirely consistent with what is known about the systems from their HST-based star-formation histories. 

From their integrated light profiles, we also measured half-light radii and \textit{V}-band magnitudes. We find that the dwarf galaxy GCs span a wide range of sizes and magnitudes, and generally occupy the same region of parameter space as other LG GCs. Curiously, the three metal-poor red-HB clusters in NGC\,6822 are also confirmed to be extended clusters (SC1, SC2 and SC4; $r_h \sim 13-17$pc). 
All of the GCs in NGC\,147 and NGC\,185, are reasonably compact ($r_h < 10$pc).

In addition to SC1, SC2 and SC4, NGC\,6822 hosts another fascinating cluster, SC7. The most luminous cluster within our sample of 26 GCs, SC7, has previously been reported to be very massive, highly elliptical, and to have a solar-scaled alpha abundance; previous work has also highlighted potential similarities between the cluster and $\omega$ Centauri, the most massive GC in the MW, and one widely thought to be the remnant core of a stripped dwarf galaxy.  We apply ellipse fitting to SC7 and find evidence for an abrupt increase in ellipticity at large radius, coincident with a small excess in light in the surface brightness profile. This behaviour is consistent with the effects of tides acting on the cluster, an interpretation that is further supported by the fact that the distortion is aligned with the direction towards NGC\,6822's centre. The presence of tidal distortion suggests that SC7 may have been orbiting within the potential well of NGC\,6822 for some time.  Chemodynamical studies of SC7 will be required to confirm the role of tides in flattening its outer isophotes and to search for evidence of a metallicity spread that would support the hypothesis that it, too, is a remnant core of an accreted dwarf galaxy. 

Our line-of-sight distance measurements for the GCs in NGC\,6822 are strongly bimodal, with the metal-poor GCs located at a mean line-of-sight distance of $536$~kpc and the more metal-rich GCs at a mean distance of $463$~kpc. This result is not yet understood, and the poorly-constrained distance to NGC\,6822 means that either group
could, in principle, be consistent with the galaxy's distance. We have highlighted possible systematic effects affecting the distance determination for GCs with low metallicities and red HBs; however, we argue that the most likely such effects would only increase the offset between the distance measurements for the metal-rich and metal-poor red-HB GCs. Therefore, we cannot rule out the possibility of a genuine line-of-sight spread in the distances of the NGC\,6822 GCs. 
Further investigation is necessary to clarify the three-dimensional spatial distribution of these clusters.

\section*{Acknowledgements}
We thank the referee for their insightful comments which improved the paper.
We warmly thank David Behrendt and Patricio Amaro Correa for their help with the early stages of this project. 
We also gratefully acknowledge discussions with Santi Cassisi, Maurizio Salaris and Anna Lisa Varri on aspects of this work. 
GM and JMH acknowledge funding from the Bell Burnell Graduate Scholarship Fund [grant numbers BB0027 and BB0015].  AMNF and CC are supported by UK Research and Innovation (UKRI) under the UK government’s Horizon Europe funding guarantee [grant number EP/Z534353/1] and by the UK Science and Technology Facilities
Council [grant number ST/Y001281/1].
This research has made use of the NASA/IPAC Extragalactic Database, which is funded by the National Aeronautics and Space Administration and operated by the California Institute of Technology.
We also made use of Astropy:\footnote{http://www.astropy.org} a community-developed core Python package and an ecosystem of tools and resources for astronomy \citep{astropy:2013, astropy:2018, astropy:2022}.

For the purpose of open access, the author has applied a Creative Commons Attribution (CC BY)
licence to any Author Accepted Manuscript version arising from this submission.

%%%%%%%%%%%%%%%%%%%%%%%%%%%%%%%%%%%%%%%%%%%%%%%%%%
\section*{Data Availability}
The raw data used in this paper are available from the Mikulski Archive for Space Telescopes (MAST) with the programme ID GO-15336. Processed data and photometric catalogues may be made available upon reasonable request.

%The inclusion of a Data Availability Statement is a requirement for articles published in MNRAS. Data Availability Statements provide a standardised format for readers to understand the availability of data underlying the research results described in the article. The statement may refer to original data generated in the course of the study or to third-party data analysed in the article. The statement should describe and provide means of access, where possible, by linking to the data or providing the required accession numbers for the relevant databases or DOIs.

%%%%%%%%%%%%%%%%%%%% REFERENCES %%%%%%%%%%%%%%%%%%

% The best way to enter references is to use BibTeX:

\bibliographystyle{mnras}
\bibliography{dwarfs}

% Alternatively you could enter them by hand, like this:
% This method is tedious and prone to error if you have lots of references
%\begin{thebibliography}{99}
%\bibitem[\protect\citeauthoryear{Author}{2012}]{Author2012}
%Author A.~N., 2013, Journal of Improbable Astronomy, 1, 1
%\bibitem[\protect\citeauthoryear{Others}{2013}]{Others2013}
%Others S., 2012, Journal of Interesting Stuff, 17, 198
%\end{thebibliography}

%%%%%%%%%%%%%%%%%%%%%%%%%%%%%%%%%%%%%%%%%%%%%%%%%%

%%%%%%%%%%%%%%%%% APPENDICES %%%%%%%%%%%%%%%%%%%%%

%\appendix

%\section{Some extra material}

%If you want to present additional material which would interrupt the flow of the main paper,
%it can be placed in an Appendix which appears after the list of references.

%%%%%%%%%%%%%%%%%%%%%%%%%%%%%%%%%%%%%%%%%%%%%%%%%%

% Don't change these lines
\bsp	% typesetting comment
\label{lastpage}
\end{document}